\documentclass[amsmath,amssymb,prl,aps,showpacs,superscriptaddress,twocolumn]{revtex4-1}
\pdfoutput=1
\usepackage{amsmath,amsfonts,amssymb,amsthm,graphics,graphicx,epsfig,bbm}
\usepackage[colorlinks=true,citecolor=blue,linkcolor=blue,urlcolor=blue]{hyperref}
\usepackage[usenames]{color}
\usepackage{graphicx}
\usepackage{subfigure}
\usepackage{amsmath}
\usepackage{epsfig}
\usepackage{dcolumn}
\usepackage{bm}
\usepackage{color}
\usepackage{epstopdf}
\usepackage{amssymb}
\usepackage{amstext}
\usepackage{latexsym}
\usepackage{hyperref}
\usepackage{amsfonts}
\usepackage{psfrag}
\usepackage{xcolor}
\usepackage[normalem]{ulem}
\usepackage{dsfont}
\usepackage{txfonts}
\usepackage{cases}
\usepackage{booktabs}
\usepackage{array}
\graphicspath{{./ImagesTopological/},{./ImagesTopologicalSuppl/}}

\newcommand{\ket}[1]{\ensuremath{\left\vert #1 \right\rangle}}
\newcommand{\bra}[1]{\ensuremath{\left\langle #1 \right\vert}}

\newcommand{\braket}[2]{\langle #1 \vert #2 \rangle}

\newcommand{\an}[2]{\ensuremath{\hat{#1}^{\protect\phantom{\dagger}}_{#2}}}
\newcommand{\cn}[2]{\ensuremath{\hat{#1}^\dagger_{#2}}}
\newcommand{\nn}[2]{\ensuremath{\hat{n}^{#1}_{#2}}}

\newcommand{\expU}[1]{\ensuremath{e^{#1}}}
\newcommand{\abs}[1]{\left|#1\right|}
\newcommand{\revA}[1]{#1}

\newcommand{\revC}[1]{#1}
\newcommand{\revD}[1]{#1}

\newcommand{\revF}[1]{#1}
\newcommand{\rev}[1]{#1}
\newcommand{\revG}[1]{#1}
\newcommand{\subfigimg}[3][,]{%
	\setbox1=\hbox{\includegraphics[#1]{#3}}% Store image in box
	\leavevmode\rlap{\usebox1}% Print image
	\rlap{\hspace*{2pt}\raisebox{\dimexpr\ht1-0.5\baselineskip}{{\bfseries \large\textsf{#2}}}}% Print label
	\phantom{\usebox1}% Insert appropriate spcing
}

\newcommand{\idg}[1]{{\bfseries #1)}}

\begin{document}
	
\title{Topological pumping in Aharonov-Bohm rings}

\author{Tobias Haug}
\email{tobias.haug@u.nus.edu}
\affiliation{Centre for Quantum Technologies, National University of Singapore,
3 Science Drive 2, Singapore 117543, Singapore}

\author{Rainer Dumke}
\affiliation{Centre for Quantum Technologies, National University of Singapore, 3 Science Drive 2, Singapore 117543, Singapore}
\affiliation{Division of Physics and Applied Physics, Nanyang Technological University, 21 Nanyang Link, Singapore 637371, Singapore}
\affiliation{MajuLab, CNRS-UNS-NUS-NTU International Joint Research Unit, UMI 3654, Singapore}
\author{Leong-Chuan Kwek}
\affiliation{Centre for Quantum Technologies, National University of Singapore,
	3 Science Drive 2, Singapore 117543, Singapore}
\affiliation{MajuLab, CNRS-UNS-NUS-NTU International Joint Research Unit, UMI 3654, Singapore}
\affiliation{Institute of Advanced Studies, Nanyang Technological University,
	60 Nanyang View, Singapore 639673, Singapore}
\affiliation{National Institute of Education, Nanyang Technological University,
	1 Nanyang Walk, Singapore 637616, Singapore}

\author{Luigi Amico}
\affiliation{Centre for Quantum Technologies, National University of Singapore,
3 Science Drive 2, Singapore 117543, Singapore}
\affiliation{MajuLab, CNRS-UNS-NUS-NTU International Joint Research Unit, UMI 3654, Singapore}
\affiliation{Dipartimento di Fisica e Astronomia, Via S. Sofia 64, 95127 Catania, Italy}
\affiliation{CNR-MATIS-IMM \&   INFN-Sezione di Catania, Via S. Sofia 64, 95127 Catania, Italy}
\affiliation{LANEF {\it 'Chaire d'excellence'}, Universit\`e Grenoble-Alpes \& CNRS, F-38000 Grenoble, France}

%\author{Rainer Dumke}
%\affiliation{Centre for Quantum Technologies, National University of Singapore, 3 Science Drive 2, Singapore 117543, Singapore}
%\affiliation{Division of Physics and Applied Physics, Nanyang Technological University, 21 Nanyang Link, Singapore 637371, Singapore}

\date{\today}

\begin{abstract}
%The progress in cold atom quantum technology enable the creation of matter wave circuits confined  in optically and magnetically generated guides  with unprecedented flexibility and control.  
%Recent achievements in this field allow  to manipulate circuits dynamically and on micrometer scales. To highlight on these technological  breakthroughs, we propose  driving in matter-wave circuits to realize topological pumping. Specifically, we study the interplay of topological pumping with the topological phase winding created by an Aharonov-Bohm flux in a bosonic ring condensate. 
Topological Thouless pumping and Aharonov-Bohm effect are both fundamental effects enabled by the topological properties of the system. Here, we study both effects together: topological pumping of interacting particles through Aharonov-Bohm rings. This system can prepare highly entangled many-particle states, transport them via topological pumping%with topological protection 
and interfere them, revealing a fractional flux quantum. The type of the generated state is revealed by non-trivial Aharonov-Bohm interference patterns that could be used for quantum sensing. The reflections induced by the interference result from transitions between topological bands. %Other types of states can be robustly transported with a band gap scaling as the square-root of the particle number.
Specific bands allow %robust 
transport with a band gap scaling as the square-root of the particle number.
Our system paves a new way for a combined system of state preparation and topological protected transport.
%We demonstrate that the Aharonov-Bohm effect in our system emerges from transitions between the topological bands. 
%Flux quantization, transmission and type of pumped states are highly dependent on interaction, number of particles and the topological band. 
%Particle-particle interaction in conjunction with topological pumping can create entangled  states in the ring which yield non-trivial Aharonov-Bohm interference patterns \revD{that could be used for quantum sensing.}
\end{abstract}
%In particular,  atomtronic circuits can be realized, with features that can be locally varied  within the lifetime of the coherent matterwave in the circuit. 
%This can be studied in a time-modulated lead-ring system pierced by a synthetic magnetic field. Atomtronics allows to study this problem with its versatile control over the potential in space %and time in cold atom system. 

%\pacs{03.65.Yz, 03.67.Lx, 42.50.Ct, 42.50.Pq}
 \maketitle

%{\it Introduction}.
%Quantum technology intertwines basic research in quantum science and technological progress
%The defining goal of quantum technology is to make fundamental research in quantum physics available for technological progress\cite{dowling2003quantum}.
\section{Introduction}
Topological matter defines  an important field in fundamental physics with far reaching implications for quantum technology:
 the correlations encoded in topological matter are  a precious resource for quantum technology; at the same time, quantum technology can be exploited to study topological matter  with unprecedented precision and control\cite{RevModPhys.89.040501,RevModPhys.89.040502,hasan2010colloquium,dowling2003quantum}.
Among the several important contributions given by David Thouless in this field, the idea of topological pumping is particularly  relevant  for quantum technology:  charge is transported through a  one dimensional system using %exploiting the topological features of the  
the topological band structure of an extended (many-body) system\cite{thouless1982quantized,thouless1983quantization}. This is realized by driving the system periodically in time while protecting the band gaps. %Specifically, the phenomenon is carried out  by deforming the system periodically in time, while protecting the band gaps.

\revG{Topological pumping has been studied experimentally in various systems, including cold atoms\cite{bloch2016,takahashi2016,lohse2018exploring}, photonic waveguides\cite{zilberberg2018photonic} and superconducting circuits\cite{tangpanitanon2016topological}. Due to interference effects, topological pumping displays very interesting features in ring-shaped networks \cite{citro2006pumping,marra2015fractional}.}
%Here, we are inspired by the new scenarios opened up by  Atomtronics: 
%as implied by the latest progress in the coherent manipulation and control of matter wave  quantum technology: 
%Here, we couple topological bands in a matter-wave circuit: 
\revG{Here, we couple topological bands in interacting Aharonov-Bohm (AB) rings attached to leads. %{\it with interaction}. 
The lead-ring interfaces act as non-linear beam-splitters. We demonstrate how interaction in such beam splitters can generate highly entangled states, %separate and transport them with {\it topological protection},
that can be revealed through interference patterns characterized by a fractional flux quantum.}

\revG{Among other  quantum technology implementations, we want to highlight Atomtronics: cold atoms matter-wave circuits \cite{seaman2007atomtronics,amico2005quantum,Amico_NJP,dumke2016roadmap} guided by laser generated fields to realize arbitrary potential configurations.}
%atomic circuits of ultra-cold atoms manipulated in micro-magnetic or laser-generated micro-optical guides . 
These systems have reduced decoherence rate due to charge neutrality, and allow one to manipulate carrier statistics and inter-atom interaction. \revG{Rectilinear circuits have been used to study quantum transport\cite{brantut2012conduction,krinner2015observation,husmann2015connecting,krinner2017two}. More complex networks can be fabricated, trapping Bose-Einstein condensates in versatile potentials. Such potentials can be changed in shape and intensity at time
	scales of tens to hundreds microseconds, and therefore
	opening the way to modify the features of the circuit in
	the course of the same experiment (typically involving
	tens of milliseconds)\cite{wright2013driving,Ryu2013,eckel2014hysteresis,amico2014superfluid,aghamalyan2015coherent,aghamalyan2013effective,Mathey_Mathey2016,haug2018readout,haug2018mesoscopic,dalibard2011colloquium,gauthier2016direct,muldoon2012control,Boshier_painting,haase2017versatile}.
% Clearly,  this design implies several advantages compared to electron-based networks, including a reduced decoherence rate due to charge neutrality of atomic currents, the ability to play with fermionic or bosonic carriers and change intrinsic parameters like the carrier-carrier interaction. By investigating quantum transport  with such new twists \cite{brantut2012conduction,krinner2015observation,husmann2015connecting,krinner2017two}, the scope of cold atoms quantum technology can be enlarged for  quantum simulation, quantum devices and quantum sensors\cite{dumke2016roadmap}.
Such remarkable advances on
the flexibility and control of cold-atoms quantum technology, in turn, has been opening up exciting possibilities for
atomtronics to study transport. %Here, we demonstrate how such new features are exploited to transport particles in atomtronic circuits through topological pumping. 
}

\revG{Previous studies on bosonic AB attached to leads have demonstrated that fundamentally new effects emerge \cite{haug2017aharonov,haug2019andreev}.
We shall see that, accordingly, topological pumping in these networks displays peculiar features as well. }
 %However, we stress that with interaction the dynamics fundamentally changes and new effects emerge: For example, it can wash out the Aharonov-Bohm effect in certain regimes \cite{haug2017aharonov,haug2018andreev}.

%We now discuss the results for different topological bands in respect to flux, ring length and particle number. The compiled results of this section is presented in Table \ref{tab:AB}.

%{\bfseries Summary of the results.}
We now summarize our results. %The setup is depicted in Fig.\ref{NAPumping} --see Eqs.\ref{ring}, \ref{sources}, \ref{driving}. 
The system is characterized by topologically distinct bands that can be controlled by changing the parameter of the driving (its phase $\phi_0$). %--see Fig.\ref{IntPumping}a,b. 
\revF{We investigate the limit where the lattice potential \revG{is sufficiently large so that particles are confined to each site during the transport.}}
%Interaction, AB oscillations and the topology of the bands couple together in a non-trivial way.  

%As a result, a  complex variety of flux quanta, types of entangled states and transmission coefficients are found:
%\revA{In particular, we understand that AB oscillations in our circuit result from transitions between the topological bands.}
%{\it {i)}}
\revG{Our device works as a non-linear interferometer, in which the source-ring and the ring-drain interfaces act as ``beam-splitters". We identify the mechanism behind  the AB effect  in topological pumped systems \revD{with interaction}: \revD{AB interference affects \revG{particle} reflections by inducing specific transitions between the topological bands. Interaction adjusts the transmission and the reflection coefficients. In addition, it can create NOON-type entanglement comprising of particles being in upper and lower arm of the ring}.} %--see Fig.\ref{IntPumping}c, \ref{NOON}.%The AB oscillations  result from transitions between topological bands.} 
%\blue{Luigi: How do you select the band you start from?? Depending on ?????}
%By choosing an initial potential configuration, we can initialize the particles in a specific band. 
%We summarize our main findings:
%{\it ii) }%Interestingly enough, 
%\revB{As a matter of fact, our device works as a {\it non-linear interferometer}, in which the source-ring and the ring-drain interfaces act as ``beam-splitters"}. 
%\revF{It generates NOON-like states \revG{made of particles being in upper and lower arm of the ring}} --see Fig.\ref{IntPumping}c, \ref{NOON}.
%Specifically: Off-resonant transfer produces entangled NOON-states, while resonant transitions produce non-entangled product type states. 
%The speed of the states preparation  is limited by the  adiabatic conditions of  the driving 
%${\log(\Omega/J)\ll-(N-1)\log(U/J)}$ 
%${\Omega/J\ll(J/U)^{N-1}}$ \cite{compagno2017noon} and  Landau-Zener transitions; the relevant energy gap depends on the curvature of the band which implies different methods of transitions (resonant and off-resonant) --see Fig.\ref{IntPumping}b.

%{\it ii) }
\revA{\revG{The speed of state preparation %and the robustness against static disorder 
\revF{in our protocol} is limited by the band gap and Landau-Zener transitions between bands.} %e.g. ${\Delta E\propto{(J/U)^{N-1}}}$ for NOON states\cite{compagno2017noon} 
\revG{The topological pumping through the two branches forming the anti-crossing is bound to be characterized by two different mechanisms: Pumping through the upper (lower) branch occurs via off-resonant (resonant) tunneling.}} %--see Fig.\ref{IntPumping}b.}}
%The anti-crossings of bands imply different methods of transitions (atoms tunnel via resonant or off-resonant transitions) --see Fig.\ref{IntPumping}b.} 
\revG{Our result shows that the transport of $N$ particles is best done in the lowest band since the band gap here scales as ${\Delta E\propto\sqrt{N}}$ (the band gap in the upper band decreases exponentially with $N$ instead).}

%{\it iii) }
The  states are transported through the ring, and interfere \revG{at the ring-drain interface; this results in a partial transmission of the particles, which is modified by the applied AB flux. }
\revG{We observe that the periodicity of the flux is reduced compared to its single particle value depending on the type of (entangled) state and particle number. This suggest that the flux quantum  $\Phi_0$ becomes a fraction of its single-particle value~\cite{leggett}.} %(see  Fig.\ref{DrainDensFlux}b)\cite{leggett}}.
%\revG{We find that the flux quantum  $\Phi_0$ becomes {\it a fraction of its single-particle value} due to entanglement: The periodicity of the transmission as a function of flux is reduced depending on the type of (entangled) state and particle number} (see  Fig.\ref{DrainDensFlux}b). 
\revC{Specifically: 
in the lowest topological band, even number of  particles are transmitted independently of the flux while  odd number of particles show AB oscillations; %--see Fig.\ref{DrainDensFlux}a;  
in the highest topological band,  AB flux periodicity changes with particle number due \revG{to the formation of NOON-like entangled states}; %--see Fig.\ref{DrainDensFlux}b;
in the central band,  different types of partial transmission and (entangled) states occur depending on the initial phase shift of the driving and the length of the ring.} 
%\revD{In particular, by changing \revF{the initial phase} the system is either transmitting or reflecting nearly all incoming particles --see Fig.\ref{Drainband0}.}
\revC{Entangled states of NOON-type \revG{of several particles} {\tiny }can be created with nearly unit fidelity.}  %--see Fig.\ref{NOON}.}

\section{Results}
\begin{figure}[htbp]
	\centering
	\subfigimg[width=0.49\textwidth]{}{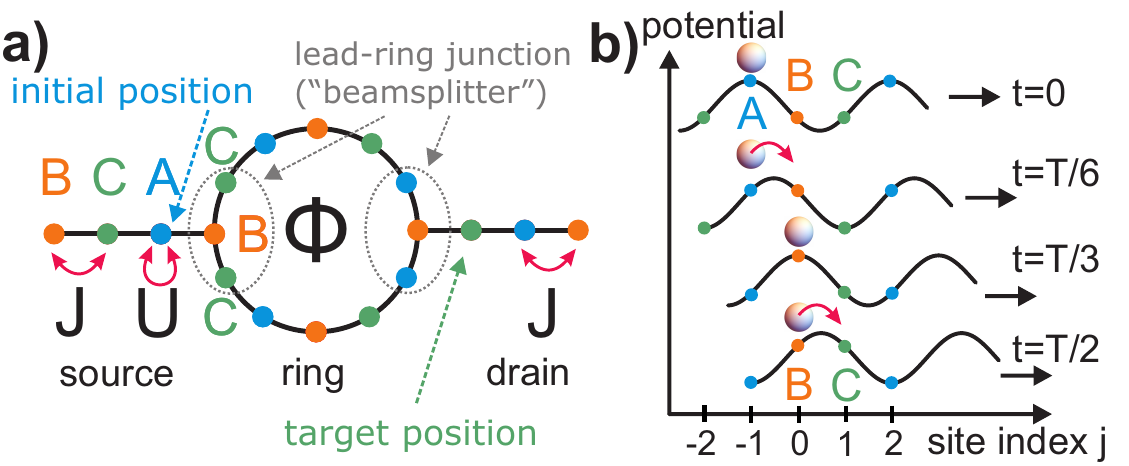}
	\caption{{\it Description of setup} \idg{a} Sketch of the ring-lead system. The dots indicate lattice sites $j$ with a space and time-dependent potential $V_j(t)=P_0\cos(2\pi/3j-\Omega t)$ with a period of three lattice sites (${\text{A:}\, j=0}$, ${\text{B:}\, j=1}$, ${\text{C:}\,j=2}$). Particles tunnel between different sites along the black lines with strength $J$. $U$ denotes the on-site interaction and $\Phi$ the flux of the ring. With topological pumping, particles can be transported from the initial position through the ring to the target drain position. The circled lead-ring junctions act as effective beam-splitters of incoming particles.
	\idg{b} Topological pumping of particles on lattice sites by adiabatic modulation of the periodic potential $V_j(t)$ with Chern number ${\mathcal{C}=-1}$. The cosine potential (black line) is varied adiabatically in time $t$ with a frequency $\Omega=2\pi/T$, where $T$ is the time of one driving period. Particles are initialized at $A$. At time ${t=0}$ no tunneling occurs due to a large potential difference to neighboring sites. At time ${t=1/6T}$, potential at $A$ and $B$ becomes degenerate, and particles are adiabatically transferred to $B$. This process continues and after one period $T$ the particles have moved by $3\mathcal{C}$ sites.}
	\label{NAPumping}
\end{figure}

{\bfseries The setup}
A sketch of the ring-lead system is presented in Fig.\ref{NAPumping}a. 
It is composed of a lattice ring attached to two leads (source and drain) symmetrically at two opposite sites of the ring loaded with $N$ particles.

\revC{One particular experimental realization is possible using a combination of experimentally demonstrated technologies for cold atom systems. Spatial light modulators and digital mirror devices (DMD) can be employed to generated arbitrary light fields in a 2D plane \cite{mcgloin2003applications,gaunt2012robust,gauthier2016direct}. With DMD, the potential can also be dynamically varied with a refresh rate of tens of $\mu$s. Similar structures can also be achieved by painting these geometries with a focused laser beam \cite{Boshier_painting}.
%Feedback algorithms can be used to further optimize the experimental realized potential. 
A confinement in a 2D plane is realized by an additional blue detuned standing wave lattice\cite{amico2014superfluid}. 
These technologies would enable the experimental realization of the proposed confining structures with feature sizes limited by the optical resolution. With these technologies, a ring lattice of tens of atoms, with a typical lattice spacing of few $\mu$m has been realized\cite{amico2014superfluid}. The phase of the confined atoms for artificial gauge fields can be manipulated by direct phase imprinting \cite{kumar2018producing} or transferred with a two photon Raman transition \cite{andersen2006quantized}.}

Superconducting cavities with microwave photons can realize Bose-Hubbard Hamiltonian, while allowing full control of the cavity potentials and even couplings in time \cite{tangpanitanon2016topological,song201710}. Ring structures and synthetic magnetic fields have been realized\cite{roushan2017chiral}.

%\blue{Ring lattice of   tens  lattice sites, with typical lattice spacing of few \,$\mu$m  generated employing Spatial Light Modulators\cite{amico2014superfluid} can be further refined with $DMD$. 
%Smooth ring-leads circuit can be  realized by playing with the optimization algorithms\cite{??}.
%$DMD$ allow a  refresh  rate of  the order of tens of \,$\mu$s to modify  the potential dynamically. Several routes  have been followed to realize  synthetic  magnetic fields\cite{dalibard2011colloquium}.  Condensates of Rb...two dimensional light sheets...Suitable  choices here  would be  shining the system with  combined gaussian and Laguerre-Gauss beams  (realizing two-photons Raman transitions)\cite{two-raman-phillips}  or by applying a suitable phase mask\cite{kumar2018producing}. }

Our model lead-ring Hamiltonian is ${\mathcal{H}=\mathcal{H}_\text{R}+\mathcal{H}_\text{S}+\mathcal{H}_\text{D}+\mathcal{H}_\text{I}+\mathcal{H}_\text{P}}$. The ring Hamiltonian is
\begin{equation}
\mathcal{H}_\text{R}=-\sum_{j=1}^{L_\text{R}}\left(J\expU{i2\pi\Phi/L}\cn{a}{j}\an{a}{j+1} + \text{H.C.}\right)+\frac{U}{2}\sum_{j=1}^{L_\text{R}}\nn{\text{a}}{j}(\nn{\text{a}}{j}-1)\;,
\label{ring}
\end{equation}
where $\an{a}{j}$ ($\cn{a}{j}$) are the annihilation (creation) operator at site $j$ in the ring, $L_\text{R}$ the number of ring sites, ${\nn{a}{j}=\cn{a}{j}\an{a}{j}}$ is the particle number operator of the ring, $J$ is the inter-site hopping, $U$ is the on-site interaction between particles and $\Phi$ is the total flux through the ring. Periodic boundary conditions are applied for the ring with ${\cn{a}{L_\text{R}}=\cn{a}{0}}$. In the following, we set ${J=1}$ and all values of $U$, $\Omega$ are given in units of $J$.
The Hamiltonian for the source (similar for drain $\mathcal{H}_\text{D}$) is given by 
\begin{equation}
\mathcal{H}_\text{S}=-\sum_{j=1}^{L_\text{S}}\left(J\cn{s}{j}\an{s}{j+1} + \text{H.C.}\right)+\frac{U}{2}\sum_{j=1}^{L_\text{S}}\nn{\text{s}}{j}(\nn{\text{s}}{j}-1)\;,
\label{sources}
\end{equation}
where $L_\text{S}$ ($L_\text{D}$) is the length of the source (drain) and $\an{s}{j}$ ($\an{d}{j}$) the source (drain) annihilation operator. We set ${L_\text{S}=L_\text{D}=1}$ in the following.
The coupling Hamiltonian between leads and ring is ${\mathcal{H}_\text{I}=-J\left(\cn{a}{0}\an{s}{0} + \cn{a}{L_\text{R}/2}\an{d}{0}+ \text{H.C.}\right)}$.
%\subsection{The protocol}\label{RingTop}
We modulate the potential landscape adiabatically to pump particles
\begin{equation}
\mathcal{H}_\text{p}(t)=P_0\sum_j \cos\left(\frac{2\pi j}{3}-\phi_0-\Omega t\right)\nn{}{j} \;,
\label{driving}
\end{equation}
with the driving frequency $\Omega$, the particle number operator $\nn{}{j}$ and phase shift $\phi_0$. The potential has a period of three sites and its arrangement in the ring-lead system is shown in Fig.\ref{NAPumping}a. 
%The arrangement of the potential components at consecutive sites are arranged in a repeating, consecutive manner from one site to the next, starting from the source lead through ring into the drain lead (see Fig.\ref{NAPumping}a). 
\revA{In the case of zero interaction and no flux,} the Hamiltonian is known to be topological nontrivial.%\revD{We define the Chern number $\mathcal{C}$
%\begin{equation}
%\mathcal{C}=\frac{1}{2\pi T}\int_{0}^{T}\int_{0}^{2\pi}
%\end{equation}
The system has three bands and non-zero Chern numbers \cite{thouless1983quantization} (top band +1 and bottom band -1 with Chern number ${\mathcal{C}=-1}$ and central band 0$^\pm$ with ${\mathcal{C}=2}$). \revA{The pumping is induced by breaking time-translational symmetry via driving. The topological properties hold true even in the interacting case \cite{tangpanitanon2016topological,nakagawa2018breakdown}. %We find that breaking time-reversal symmetry with flux $\Phi$ generates transitions into other bands, while leaving the topological properties of the bands intact. 
} Similar systems have been studied in \cite{tangpanitanon2016topological,he2018topology}. After one period of adiabatic time evolution ${T=2\pi/\Omega}$, particles move by $3\mathcal{C}$ sites (see Fig.\ref{NAPumping}b). 
The protocol is as follows: Initially, a Fock state with $N$ particles at a single site in the source lead is prepared. Here, we initialize the particles at the first source lead site that directly neighbors the ring and tune $\phi_0$ to select the band (see Fig.\ref{NAPumping}a).
%Depending on the pumped band, different types of states are created, which also react differently to an applied phase shift, e.g. by an Aharonov-Bohm flux. 
%The dynamics of this band depends on the initial phase shift ${\phi_0^{+}=\pi/2}$ (denoted as 0$^+$)  and ${\phi_0^{-}=-\pi/2}$ (denoted as 0$^-$).

In the following, we investigate positive interaction ${U>0}$ without loosing generality. For ${U<0}$, simply switch the results of band +1 with -1, and 0$^+$ with 0$^-$.
\rev{We operate in the limit of ${P_0\gg J,U}$, such that the eigenstates are strongly localized within single sites. Thus, tunneling between neighboring sites is suppressed unless close to a resonance (since they have in general a widely different local potential energy), as well as effective tunneling across three sites (to the nearest site with the same potential). The effective tunneling dynamics can be understood in this limit by including the nearest neighboring site only.} %To fulfill the adiabatic criterion we demand as well ${\Omega\ll J,U,P_0}$.%We assume positive interaction $U$ (changing sign of $U$ is equivalent to flipping sign of $P_0$).
We define the transmitted density as the expectation value of the number of particles in the drain at the time the particles arrive in the drain (see final position in Fig.\ref{NAPumping}a). 
%Particles that are initially reflected in the ring after the first iteration may reach the drain at a later point by making several round trips inside the ring, however in the scope 
%of this paper we do not consider these further contributions. 

{\bfseries Non-interacting topological pumping}
%Now, we study the  topological pumping through the ring-lead system.
%
\begin{table*}[htbp]
	\centering
	\begin{center}
%		\begin{tabular}{cc|c|c|c|c|c|cc|cc|c|c} % <-- Alignments: 1st column left, 2nd middle and 3rd right, with vertical lines in between
%			&\textbf{band}&ring length  & Chern &$\phi_0$& AB period  & parity& \multicolumn{2}{c}{transmission: $N$ even}& \multicolumn{2}{|c|}{transmission: $N$ odd}&state in ring&band gap\\
%			& ${U>0}$&$L_\text{R}$ & number& &$\Phi_0$ &effect &${\Phi=0}$& ${\Phi=\Phi_0}$& ${\Phi=0}$& ${\Phi=\Phi_0}$&&$\Delta E$\\
%			\hline
%			&+1 & $2n$&-1& $0$& $1/N$ &no & $N$ &$N-1$&$N$ & $N-1$&entangled&$J^N/U^{N-1}$\\
%			&0$^+$ &$4n+2$&2&$\pi/2$ & $1/N$ &no & $N$ &$N-1$ & $N$ &$N-1$&entangled&$J^N/U^{N-1}$\\
%			&0$^+$ &$4n$&2&$\pi/2$ & $1/N$ &yes &$0$& $1$ & $1$ & $0$&entangled&$J^N/U^{N-1}$\\
%			&0$^-$ &$4n$&2&$-\pi/2$ & $1$  &yes& $0$ & $0$ & $1$ & $0$&product&$J^N/U^{N-1}$\\
%			&0$^-$ &$4n+2$&2&$-\pi/2$ & $1$  &yes& $N$ & $N$ &$N$ & $N-1$&product&$J^N/U^{N-1}$\\
%			&-1 &$2n$& -1& $\pi$  &$1$ &yes &  $N$ & $N$ &$N$ & $N-1$&product&$2\sqrt{N}J$\\
%			\hline
%			&${U=0}$ all bands & $2n$& & &  $1$ &no & $N$ &$0$&$N$ & $0$&superposition&$2J$\\
%		\end{tabular}
			\begin{tabular}{cc|c|c|c|c|c|c|c|c|c} % <-- Alignments: 1st column left, 2nd middle and 3rd right, with vertical lines in between
		&band&ring length  & transmission $N$ even& transmission $N$ odd& Chern &$\phi_0$& AB period  & parity&state in ring&band gap\\
		& ${U>0}$&$L_\text{R}$ & $T_\text{even}(\Phi,N)$&$T_\text{odd}(\Phi,N)$&number& &$\Phi_0$ &effect &&$\Delta E$\\
		\hline
		&+1 & $2n$ & ${N-1+\cos^2(\pi\Phi N)}$ & $N-1+\cos^2(\pi\Phi N)$&-1& $0$& $1/N$ &no&NOON type&$J^N/U^{N-1}$\\
		&0$^+$ &$4n+2$ & $N-1+\cos^2(\pi\Phi N)$ & $N-1+\cos^2(\pi\Phi N)$&2&$\pi/2$ & $1/N$ &no&NOON type&$J^N/U^{N-1}$\\
		&0$^+$ &$4n$ &$\sin^2(\pi\Phi N)$& $\cos^2(\pi\Phi N)$&2&$\pi/2$ & $1/N$ &yes&NOON type&$J^N/U^{N-1}$\\
		&0$^-$ &$4n$& $0$ & $\cos^2(\pi\Phi)$&2&$-\pi/2$ & $1$  &yes&varies&$J^N/U^{N-1}$\\
		&0$^-$ &$4n+2$& $N$ &$N-1 +\cos^2(\pi\Phi)$&2&$-\pi/2$ & $1$  &yes&varies&$J^N/U^{N-1}$\\
		&-1 &$2n$ &  $N$ &$N-1 +\cos^2(\pi\Phi)$& -1& $\pi$  &$1$ &yes&varies&$2\sqrt{N}J$\\
		\hline
		&${U=0}$ all bands & $2n$& $N\cos^2(\pi\Phi)$&$N\cos^2(\pi\Phi)$& &&  $1$ &no &superposition&$2J$\\
	\end{tabular}
		\caption{{\it Overview of results} AB period (flux quantum) $\Phi_0$ and number of particles transmitted $T(\Phi,N)$ after pumping through an interacting AB ring with flux $\Phi$ in the limit of strong localizing potential $P_0$. Results depend on pumped band, ring length, parity of particle number $N$ and interaction (here ${U>0}$). Reflection given by ${R=N-T}$. Bands are visualized in Fig.\ref{IntPumping}a.%+1 indicates the band with highest energy, -1 the lowest energy band (both Chern number ${\mathcal{C}=-1}$). 0$^+$ and 0$^-$ is the central band with ${\mathcal{C}=2}$, with different initial phase shifts $\phi_0$ in the potential. 
		For ${U<0}$, exchange the band indices $+\leftrightarrow-$ (e.g. for ${U<0}$ band +1 behaves like band -1 for ${U>0}$). %The state in the ring column shows the type of state for indistinguishable and two distinguishable particles. The transmission coefficient for two distinguishable particles is identical to the $N$ even column.
		}\label{tab:AB}
	\end{center}
	%\label{results}
\end{table*}
\begin{figure}[htbp]
	\centering
	\subfigimg[width=0.49\textwidth]{}{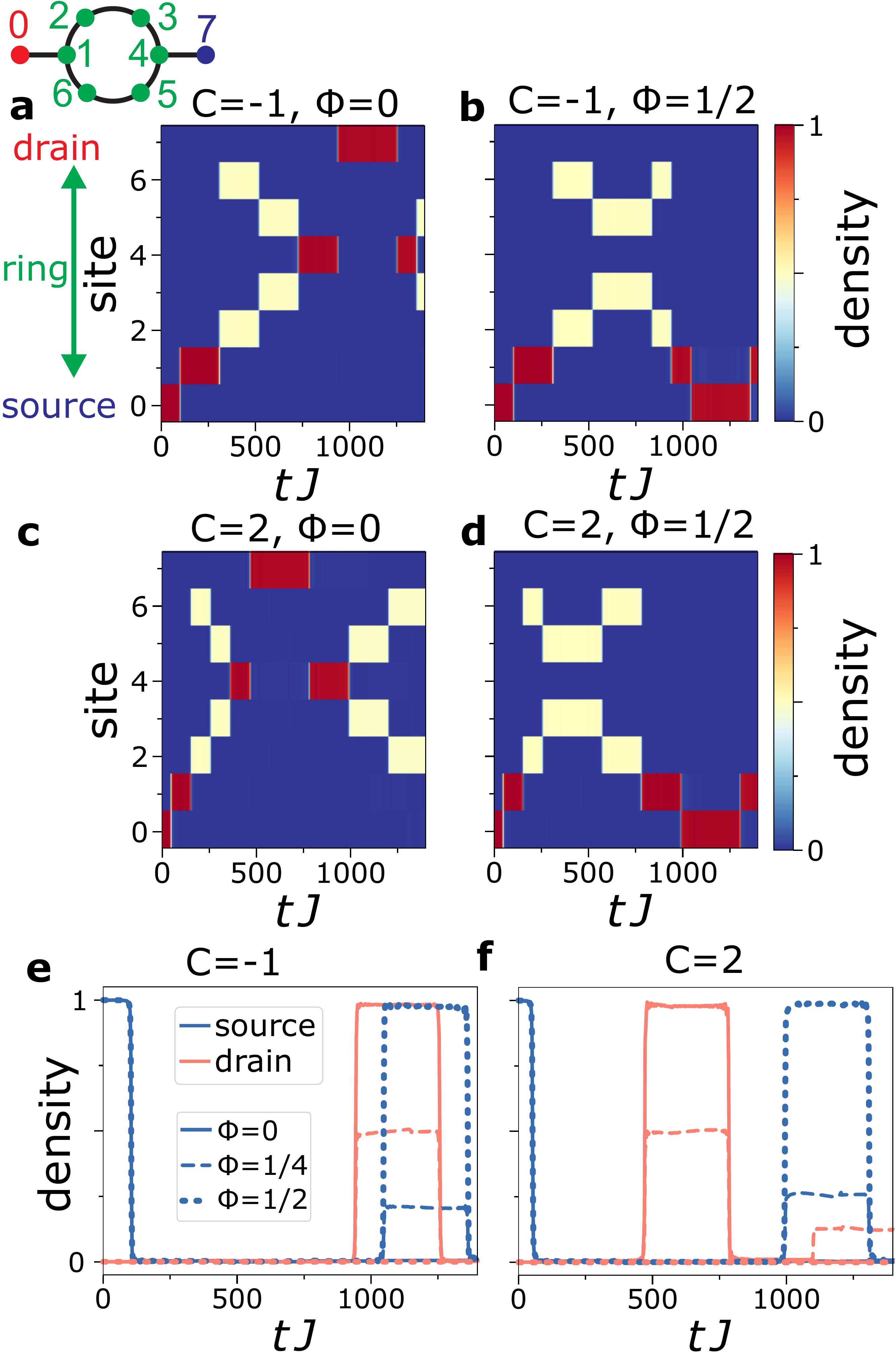}
	\caption{{\it Pumping for non-interacting ring} Time evolution of topological pumping of non-interacting particles in the ring-lead system with ring length ${L_\text{R}=6}$, ${N=1}$ particles and potential strength ${P_0=60J}$. Initial band has \idg{a,b} Chern number ${\mathcal{C}=-1}$ (driving frequency ${\Omega=0.01J}$, phase shift ${{\phi_0=0}}$) and \idg{c,d} Chern number ${\mathcal{C}=2}$ ($\Omega=-0.01J$, ${\phi_0=\pi/2}$). \idg{a,c} show flux {${\Phi=0}$}, \idg{b,d} shows flux {$\Phi=1/2$}. 
	Flux ${\Phi=1/2}$ causes total reflections; the particles are transferred to a band with Chern number of opposite sign and different velocity (\idg{b} initially in ${\mathcal{C}=-1}$, reflected particle in ${\mathcal{C}=2}$).
	Y-axis depicts ring-lead system sites, with site 0: source, site 1-6: ring,	site 7: drain. Sketch of ring is shown on the top left. %\revF{(site 1 is connected via tunneling to source, site 2 and 6 are connected to site 1, site 4 is connected to drain,  )}
  \idg{e,f} density in source and drain against time for different values of flux and Chern number \idg{e} $\mathcal{C}=-1$ and \idg{f} $\mathcal{C}=2$. 
	}
	\label{RingNAToplogicalPumping}
\end{figure}
\revD{First, we discuss the single-particle dynamics for the ring-lead system. The particles are prepared in the source at a single site as shown in Fig.\ref{NAPumping}a. %We choose ${\phi=\pi}$, which corresponds to the lowest band with Chern number ${\mathcal{C}=-1}$. 
The dynamics of the non-interacting case is plotted in Fig.\ref{RingNAToplogicalPumping}. We investigate two cases: For bands $\pm 1$ with Chern number ${\mathcal{C}=-1}$, and band 0 with ${\mathcal{C}=2}$. We highlight that the dynamics both bands $\pm 1$ are identical. We now evolve the system under the time-dependent Hamiltonian. The particles are transported towards the ring via adiabatic driving. The speed depends on the Chern number of the respective band. At the ring, the path splits into two ways: The upper and lower part of the ring. Here, the particles split into a superposition state, going along both paths at the same time. At the end of the ring, the two paths merge and interfere. We can control this interference using the artificial magnetic flux $\Phi$. Depending on $\Phi$, we observe constructive or destructive interference. For constructive interference, particles leave the ring and move into the drain. For destructive interference at ${\Phi=1/2}$, particles are reflected. In this case, we observe that the particles move back the way they came at a different speed. If the particles are initially in a band with Chern number ${\mathcal{C}=-1}$, the reflection occurs via the band with ${\mathcal{C}=2}$, and vice versa. 
From this, we can understand the mechanism how reflections in topological pumped AB rings arise: Reflections caused by destructive AB interference \rev{arise by} transferring particles to bands with a Chern number of opposite sign.}
\revC{However, we find that these AB reflections occur independently of the driving frequency, indicating that they  are distinct from Landau-Zener transitions.}
\revD{We can understand the reflection and transmission by looking at the dynamics at a reduced lead-ring junction consisting only of three sites: two sites of the ring ($\ket{C_1}$ and $\ket{C_2}$) connected via hopping to a single drain site ($\ket{B}$). This reduced three site system is a three level system for zero interaction, akin to the well known $\Lambda$-system. One of its eigenstates is a dark eigenstate ${\ket{D}=\frac{1}{\sqrt{2}}(\ket{C_1}-\ket{C_2})}$, that always has zero amplitude at the drain site for any value of the driving parameter $\phi(t)$. Conversely, there is also a bright eigenstate that will adiabatically tunnel from the two input sites over to the drain site. Particles can only tunnel over if they are in a bright eigenstate, and are prohibited from tunneling if they are in the dark state. We calculate the dynamics of an incoming superposition state $\ket{\Psi_\text{in}}=1/\sqrt{2}(\ket{C_1}+\expU{i2\pi\Phi}\ket{C_2})$, which is initialized in the two input sites and then adiabatically pumped. The transmission probability is given by the overlap with the bright state, and the reflection probability (the probability of not reaching the drain site) is the overlap of the incoming superposition state with dark eigenstate ${R=\abs{\braket{\Psi_\text{in}(\Phi)}{\Psi_\text{dark}}}^2}$. The overlap depends on the phase of the superposition state, which is controlled by flux $\Phi$. The reflection probability of this three level system is ${R=\sin^2(\pi\Phi)}$ and the transmission probability ${T=1-R=\cos^2(\pi\Phi)}$.
}
For a single particle or ${U=0}$ the flux dependence and smallest energy gap ${\Delta E=2J}$ is the same for all bands. The speed of pumped particles solely depends on the Chern number of its band.  %We see the standard AB effect with full transmission through the ring for zero flux, and total reflection due to destructive interference at half-flux. Remarkably, reflections occur by transitions into bands with Chern number of opposite sign, such that the particle moves then into the reverse direction. 
%\revD{The reflection and transmission is the result of interference of the two paths of the ring (indicating particle in upper (lower) part of ring $\ket{\cap}$ ($\ket{\cup}$). %We can easily derive the flux dependence for a single particle.}
% The transmission is ${T=\abs{1/\sqrt{2}(\ket{\cap}+\expU{i2\pi\Phi}\ket{\cup})}^2}=\cos(2\pi\Phi)$ and ${R=1-T}$.}
%\revC{However, they are distinct from Landau-Zener transitions as they occur for even very slow driving frequency.}

\begin{figure*}[htbp]
	\centering
	\subfigimg[width=0.95\textwidth]{}{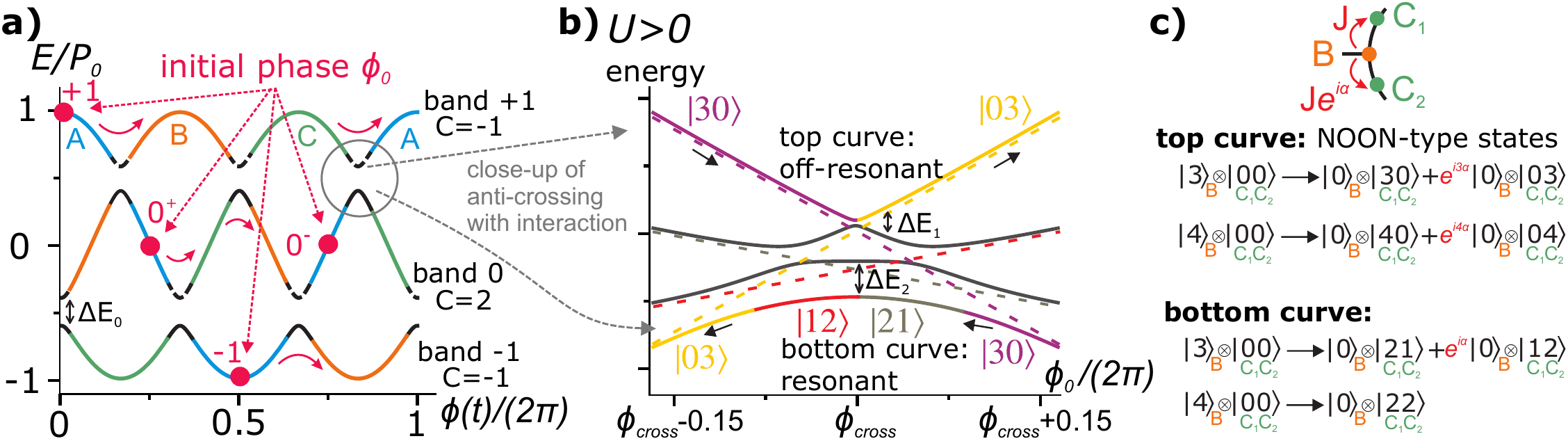}
	\caption{{\it Pumping mechanism} \idg{a} Energy spectrum of three bands for topological pumping in a lattice, with potential ${V_j(t)=P_0\cos(2\pi/3j-\phi(t))}$, ${\phi(t)=\phi_0+\Omega t}$ with a period of three sites (${\text{A:}\, j=0}$, ${\text{B:}\, j=1}$, ${\text{C:}\,j=2}$).  Particles are initialized at site $A$. %Site $A$ corresponds to different initial phase shifts $\phi_0$ for each band. 
		By adiabatically changing phase $\phi(t)$  particles follow the band, moving along $A$-$B$-$C$ across the system. %For band 0 (Chern number ${\mathcal{C}=2}$), two different initial $\phi_0$ are possible, with different dynamics. 
		The energy gap for interaction ${U=0}$ is ${\Delta E_0=2J}$ for all bands.
		\idg{b} \revC{Close-up of the interacting many-body eigenlevel structure (for repulsive interaction ${U>0}$) of \rev{any band anti-crossing (at exemplary circled area in {\bfseries a})}. We show the eigenlevel structure of two neighboring sites $j$ and ${j+1}$ as solid lines (decoupled system ${J=0}$ as dashed lines). $\ket{30}$ denotes a state with 3 particles at site $j$ and 0 particles at site ${j+1}$. Band +1  follows top curve of the anti-crossing, while band -1 follows bottom curve. Band 0 switches at every anti-crossing between top or bottom curve (see {\bfseries a}).% \rev{(not present in non-interacting picture).} 
		For the top curve of anti-crossing, the energy gap to the next eigenstate is scaling as ${\Delta E_1\propto J^N/U^{N-1}}$ for ${U\gg J}$. The transport occurs due to off-resonant coupling of the final states ($\ket{30}$ and $\ket{03}$ for ${N=3}$) via off-resonant intermediate states ($\ket{21},\ket{12}$) which are barely occupied. 
		For the bottom curve, the energy gap is ${\Delta E_2=2\sqrt{N}J}$ for ${U\gg J}$. Transport occurs via resonant transitions to intermediate states (${\ket{21},\ket{12}}$) which are significantly occupied. 
			%To understand why top and bottom level acquire different dynamics with interaction, we plot eigenlevels of a decoupled system ${J=0}$ as dashed lines. The intermediate levels ($\ket{21},\ket{12}$) are shifted downwards by interactions, thus coupling more strongly with the bottom level and breaking the symmetry between top and bottom levels.
		}
		% states are coupled resonantly via the transitions ${\ket{30}\rightarrow\ket{21}\rightarrow\ket{12}\rightarrow\ket{03}}$. 
		%and the intermediate states  are significantly occupied during the transition. 
		%For ${U<0}$, the behavior of the bands is interchanged.
		\idg{c} \rev{Pumping through the lead-ring junction} (consisting of a site $\text{B}$ connected to two sites $\text{C}_1$ and  $\text{C}_2$). \revG{This three site system effectively acts as a non-linear beam-splitter. Adiabatic driving} of the two types of transitions will produce different final states. Phase of resulting state depends on the complex tunneling strength $Je^{i\alpha}$ due to the flux $\Phi$. \rev{The top curve of the anti-crossing} gives entangled states, which pick up a phase $\propto N$ due to the NOON-type entanglement. \rev{Bottom curve yields states that depend on the parity of the particle number $N$, which either pick up a phase factor (odd $N$) or none (even $N$).} }
	\label{IntPumping}
\end{figure*}

{\bfseries Interacting topological pumping}
\revD{We now turn to the case of non-zero interaction. While topological pumping was defined above for non-interacting systems, we find the same Chern numbers (how many sites particles travel in a band per period) even for the interacting system in our configuration. 
%For our pumping scheme, which consists of shifting the potential in space, pumping should be possible for any amount of interaction\cite{nakagawa2018breakdown}. 
\rev{We believe our approach does work because our protocol can pump bound states of $N$ particles (instead of bare particles), which retain the topological properties of the non-interacting particles.}%\revF{We conjecture the following e pumping with interaction is possible: Our setup pumps effective bound states of $N$ particles. They behave effectively as free particles with $N$ times mass of the original particles, but keep the same topological properties.}%This may be related to the formation of effective bound states with interaction, which behave similar to the non-interacting case.}
}
%We find that for increasing ${\abs{U}>\Omega}$ the standard AB effect \revF{gradually} disappears, and instead we observe a many-body AB effect. 
\revF{For increasing interaction ${\abs{U}}$, many-body effects come gradually into play. Once the interaction $U>\Omega$ is larger than the driving speed $\Omega$, the energy splitting between different many-body states is large enough such that it is not washed out by the driving. In this regime, we observe a many-body AB effect.}

Then, the transmission and reflection through the ring substantially depends on the band and the sign of $U$. The pumping mechanism with interaction is described in Fig.\ref{IntPumping}. Pumping for a non-interacting system is described in Fig.\ref{IntPumping}a. Particles tunnel from one site to the next at the anti-crossings of the band (see Fig.\ref{IntPumping}b). With interaction, there is an asymmetry in the pumping mechanism for top and bottom level of the anti-crossing. Interaction pushes most of the uncoupled many-body levels downwards and closer together in energy, while the top two levels are further detached from the other levels. Thus, with nearest-neighbor coupling, the bottom level of the anti-crossing transports states resonantly via intermediate states, while the top level transport occurs off-resonantly without occupying those intermediate states. The top band has an exponentially suppressed gap with particle number, while for bottom band it increases with particle number.
Transmission and reflection for interacting particles are governed by this asymmetry at the ring-lead interface (see Fig.\ref{IntPumping}c). \revD{The many-body AB effect changes transmission and reflection by at most one particle, e.g. for band +1 in between $N$ to ${N-1}$ particles are transmitted depending on flux. The functional dependence is similar to the non-interaction AB effect, e.g. for band +1 it is ${T=N-1+\cos^2(\pi\Phi N)}$, where $N$ is the particle number. Note that in this case the interaction reduces the periodicity with flux compared to the non-interacting case, with a flux quantum of ${\Phi_0=1/N}$. A full list is shown in Table \ref{tab:AB}.}
\revA{Additionally, the energy gap becomes dependent on band, interaction and particle number.} \revD{In particular, the symmetry of bands $\pm1$ is broken, and they behave differently. Additionally, band $0$ changes its behavior depending on the initial phase $\phi_0$.}
%The change of the drain density with interaction is shown in Fig.\ref{DrainDensInt}. In this regime, we see oscillations which stem from non-adiabatic transitions as the driving $\Omega$ is on the same order as the energy gap induced by $U$.
%
%The resulting density in the drain against flux is plotted in the yellow short-long dashed curve of Fig.\ref{DrainDensFlux}a,b.

\begin{figure}[htbp]
	\centering
	%\subfigimg[width=0.24\textwidth]{a}{density0timeFy0s1m12L6N3J1U-1g1f0_0w0p40o0_01s40000n1data.pdf}\hfill
	%\subfigimg[width=0.24\textwidth]{b}{density1timeFy0s1m12L6N3J1U-1g1f0_0w0p40o0_01s40000n1data.pdf}\\
	%\subfigimg[width=0.24\textwidth]{b}{density2timeFy0s1m12L6N3J1U-1g1f0_0w0p40o0_01s40000n1data.pdf}\\
	\subfigimg[width=0.49\textwidth]{}{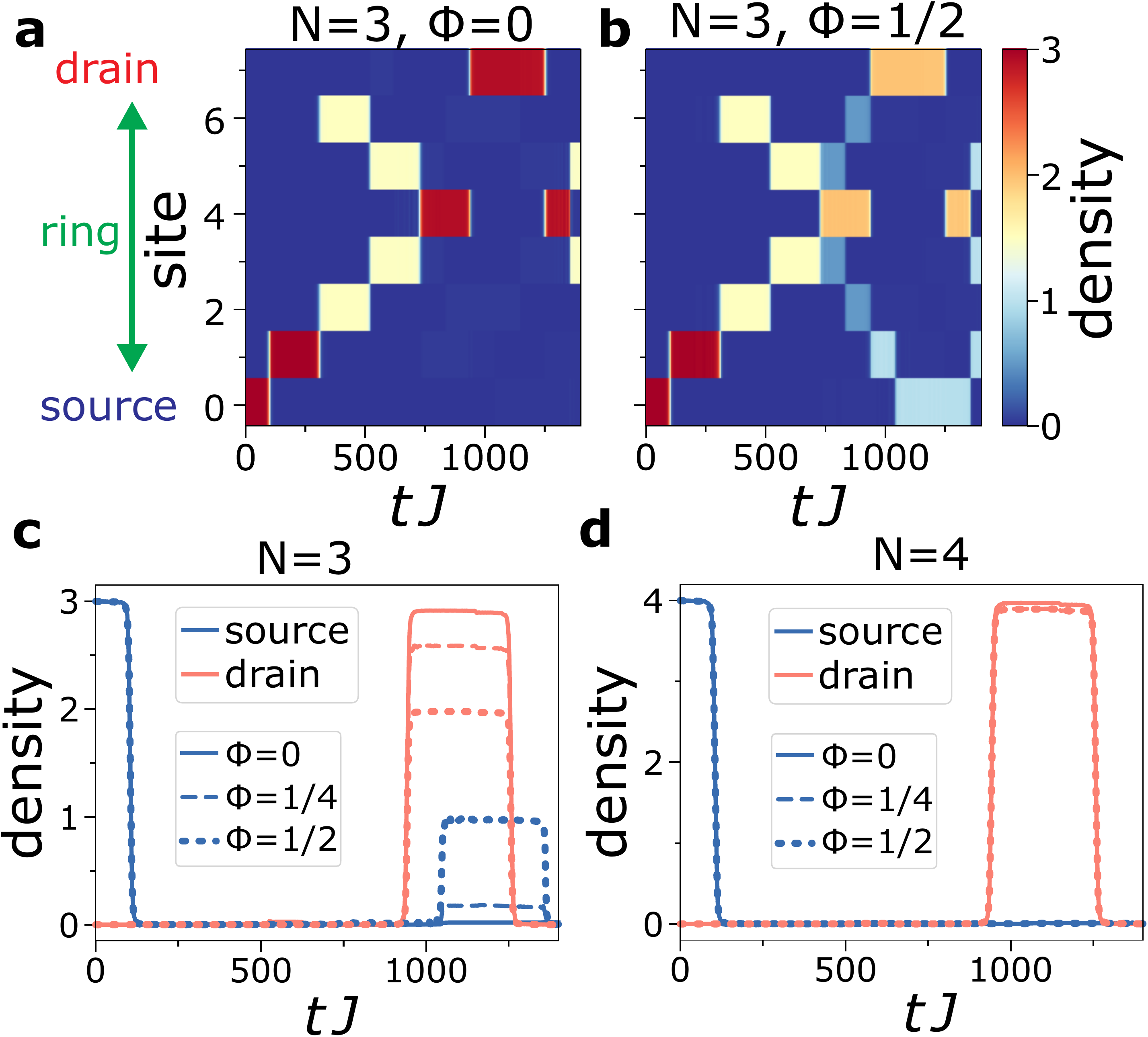}
	%\subfigimg[width=0.46\textwidth]{}{densityFulltimeFy0s1m12L6N3J1U-1g1f0_0w0p40o0_01s40000n1data.pdf}
	%\subfigimg[width=0.24\textwidth]{c}{densitySD1DtimeFy0s1m12L6N3J1U-1g1f0_0w0p40o0_01s40000n1data.pdf}\hfill
	%\subfigimg[width=0.24\textwidth]{d}{densitySD1DtimeFy0s1m12L6N4J1U-1g1K1f0_0w0p40_0o0_01s40000n0data.pdf}
	%\subfigimg[width=0.24\textwidth]{f}{densitySD1DtimeFy0s1m12L6N5J1U-1g1K1f0_0w0p40_0o0_01s40000n0data.pdf}
	\caption{{\it Pumping for interacting ring} Time evolution of topological pumping of interacting particles with ring length ${L_\text{R}=6}$, interaction ${U/J=1}$, driving frequency $\Omega=0.01J$ and potential strength ${P_0=40J}$ in band -1. \idg{a,b} Density against time and sites for ${N=3}$ particles, \idg{a} flux ${\Phi=0}$ and %\idg{b} ${\Phi=1/4}$, 
	\idg{b} ${\Phi=1/2}$. Site 0: source, site 1-6: ring, site 7: drain  \idg{c-d} density in source and drain for %(solid: ${\Phi=0}$, dashed: ${\Phi=1/4}$, dots: ${\Phi=1/2}$) 
	particle numbers \idg{c} ${N=3}$, \idg{d} ${N=4}$.
	In interacting system, partial transmission and reflection occurs: For this band -1 and ${\Phi=1/2}$, $N-1$ particles are transmitted and 1 reflected.%, \idg{f} ${N=5}$
	%Odd number of particles: drain density at ${tJ=1100}$ changes with flux: for ${\Phi=0}$) $N$ particles are transmitted, while for (${\Phi=1/2}$) ${N-1}$ particles are transmitted and 1 reflected. The reflected particle is transported back to the source via band 0 (${\mathcal{C}=2}$) at twice the speed. Even particle number: all particles are transfered to the drain independent of flux. 
}
	\label{RingToplogicalPumping}
\end{figure}
\begin{figure}[htbp]
	\centering
	%\subfigimg[width=0.4\textwidth]{}{Dynamicsband1.pdf}
	\subfigimg[width=0.49\textwidth]{}{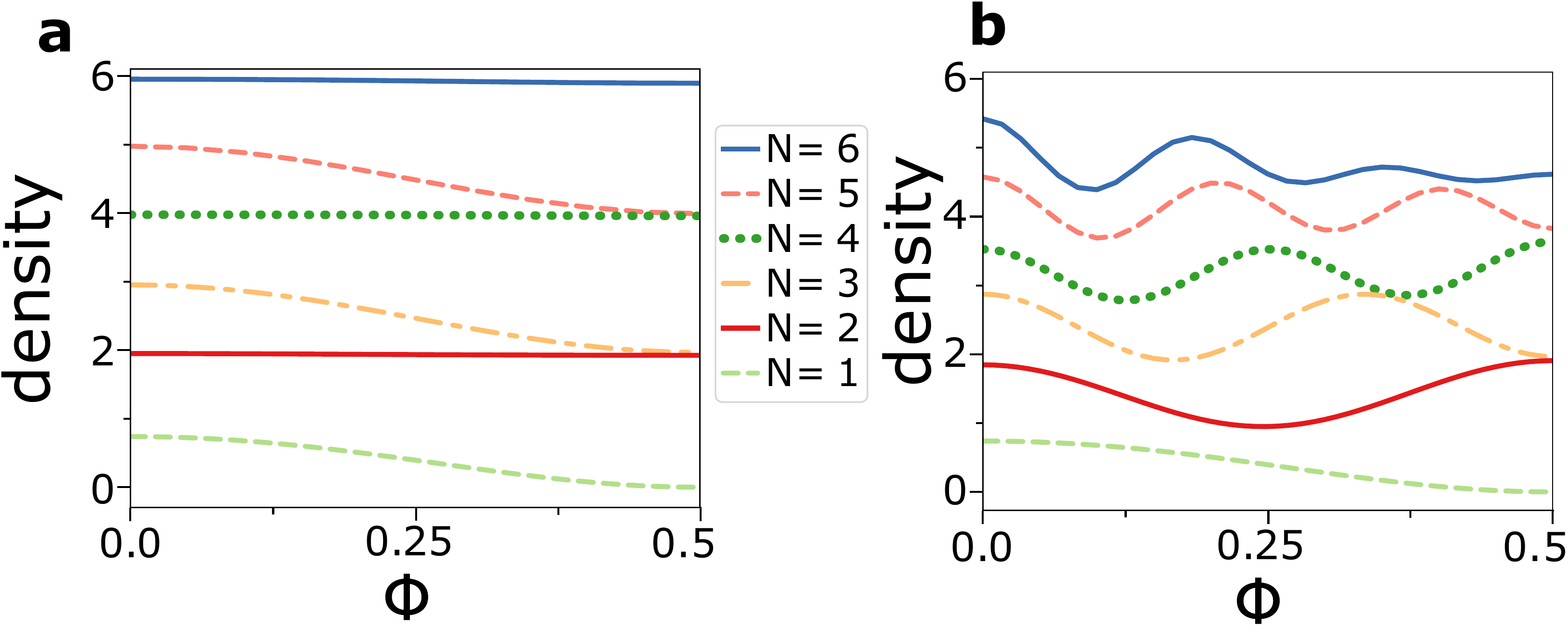}\hfill
	%\subfigimg[width=0.24\textwidth]{a}{densityParam1DFluxtimeFy0s1m12L8N4n0J1U-0_5u0_0g1K1f0_0w0p60o0_01s10000.pdf}\hfill
	%\subfigimg[width=0.24\textwidth]{b}{densityParam1DFluxtimeFy0s1m12L8N4n0J1U0_1u0_0g1K1f0_0w0p60o0_01s10000.pdf}
	\caption{{\it Transmission for top and bottom band} Density pumped into the drain against flux $\Phi$ for \idg{a} band -1, ${U=0.5}$ \idg{b} band +1 ${U=0.1}$
	\idg{a} band -1: $\Phi$ dependence changes with even/odd parity of particle number $N$. \idg{b} band +1: pumped entangled states show flux periodicity $\Phi_0\propto 1/N$ in the transmitted density.
	%($U<0$ has same dynamics as band -1 with $U>0$). 
	Drain density measured at time ${tJ=1260}$. Pumping with driving frequency ${\Omega=0.01J}$, ring length ${L_\text{R}=8}$, and potential strength ${P_0=60J}$.}
		%Density pumped through the ring into the drain plotted against flux, with ring length $L=8$ and initial band +1 ($U<0$ has same dynamics as band -1 with $U>0$). Drain density taken at ${tJ=1260}$, ${\Omega=0.01J}$ and ${P_0=60J}$.}
	%Density pumped through the ring into the drain plotted against flux, with ring length $L=6$ and initial band -1. Drain density taken at ${tJ=1100}$, ${\Omega=0.01J}$ and ${P_0=60J}$. \idg{a}  for different number of particles and ${U=1}$. For odd, Aharonov-Bohm effect changes drain density by one particle, while for even flux has no effect.
	%\idg{b} Flux dependence for particle number and different bands (band -1 ${U=0.5}$, band +1 ${U=0.1}$).}
	\label{DrainDensFlux}
\end{figure}

{\bfseries Lower band -1}
Here, we choose initial phase shift ${\phi_0=\pi}$.
%We study the dynamics three interacting bosons in the ring-lead system in Fig.\ref{RingToplogicalPumping}. 
In this case,  the avoided crossing  is approached from below. % (the bottom curve of Fig.\ref{IntPumping}b). 
The particle transfer from one site to the next  is found to happen via resonant transitions to intermediate many-body states when these states become resonant. For ${U\gg J}$, the energy gap is independent of $U$: ${\Delta E=2\sqrt{N}J}$. %This shows that the pumping becomes {\it more robust with increasing particle number $N$} (see supplementary material).
\revG{This shows that the pumping can be driven at higher frequency with increasing particle number $N$ (see supplementary materials).}
%For this type of resonant coupling, the energy gap %${\Delta E(x)=P_0-P_1 +U(N-2x-1)}$ 
%to other non-resonant states increases with $U$, thus suppressing  the non-adiabatic Landau-Zener transitions for larger $U$.
%Because of such features, the density increases as result of sequential tunneling $\ket{N-x|x} \rightarrow\ket{N-x-1|x+1}$, with the notation $\ket{N_\cap|N_\cup}$, being $N_\cap$ and $N_\cup$  the number of particles in the upper/lower arm of the ring respectively. 
%e.g. for a total particle number $N$ and $x$ particles already tunneled over, the state  will change into $\ket{N-x-1|x+1}$ while driving. The energy difference of two connected 
%many-body states $\ket{N-x|x}$ and $\ket{N-x-1|x+1}$ is . During the driving, one many-body state after the other become resonant, and 
%increase the number of tunneled particles $x$ sequentially. 
In this regime, we find that the pumping is dependent on the parity of the particle number $N$. The dynamics for different particles numbers is plotted in Fig.\ref{RingToplogicalPumping}. For ${N=2n+1}$ we find a fractional transmission with flux: For zero flux, all particles are transported to the drain (uppermost site) (Fig.\ref{RingToplogicalPumping}a). However, for half-flux, one particle is reflected and the rest is transmitted  (Fig.\ref{RingToplogicalPumping}b). The back-reflection occurs at twice the speed via the central band with Chern number ${\mathcal{C}=2}$. The density in the drain for odd $N$ is plotted in Fig.\ref{RingToplogicalPumping}c. In contrast, for even ${N=2n}$  all the particles reach the drain independently of the flux (Fig.\ref{RingToplogicalPumping}d). %The effect traces back to the different structure of the many-body states that are generated in the ring by the driving. 
To understand the parity effect, we investigate the type of Fock states generated at the ring-lead junction. We denote the wavefunction of a single particle localized in the upper half of the ring as $\ket{\cap}$, and $\ket{\cup}$ in the lower half of the ring. \rev{The Fock states are generated by the resonant tunneling process at the ring-lead junction, where the path splits into two directions. Here, particles tunnel one after the other. The process favours the state with the lowest interaction energy; thus, the system tries to achieve a state with an equal number of particles in each arm. For ${N=2n}$, such a 'balanced' state can be reached; the ring state is ${\ket{\Psi_{2n}}=(\ket{\cap}\otimes\ket{\cup})^n}$, which does not pick up any AB phase and thus causes no interference.  
For  ${N=2n+1}$, the state is always un-balanced, and therefore the extra-particle can sustain AB effect. In this case,  the state  of the ring  has the form ${{\ket{\Psi_{2n+1}}=(\ket{\cap}\otimes\ket{\cup}})^n\otimes(\ket{\cap}+\expU{i2\pi\Phi}\ket{\cup})}$;  the last part of the wavefunction reacts to flux $\Phi$ and can interfere.} %The flux quantum is ${\Phi_0=1}$.
The resulting density pumped into the drain is shown in Fig.\ref{DrainDensFlux}a.
Incidentally, we  note that a setup with two independent particle species (e.g. two internal states) can provide entangled Bell states (see supplementary materials).  

{\bfseries Upper band +1}
This band can create highly entangled states. We choose ${\phi_0=0}$ so that the avoided crossing is approached from above. % -- Fig.\ref{IntPumping}b. 
\rev{The density transmitted to the drain  is shown in Fig.\ref{DrainDensFlux}b. In this case, we find that tunneling between neighboring sites  occurs via effective tunneling between  nearly resonant states that are not connected directly via the hopping term. Here, the transitions happens via intermediate off-resonant processes.}  (e.g. for ${N=3}$ and $\ket{N_jN_{j+1}}$ denoting particles at neighboring sites $j$: when states $\ket{30}$ and $\ket{03}$ become resonant, they are off-resonantly coupled via the weakly occupied states $\ket{21}$ and $\ket{12}$).  
%The condition for this driving to work is ${\abs{P_0}\gg \abs{U} < \abs{J}}$. 
\revF{The effective coupling between the final states can be calculated with the Schrieffer-Wolff transformation using above reasoning\cite{tangpanitanon2016topological}. Then, the energy gap between the bands is $\Delta E\propto J^{N}/U^{N-1}$ (\cite{compagno2017noon}, also see supplementary materials).} The gap decreases sharply with increasing interaction $U$. Thus, the pumping is most efficient in the regime ${\abs{U}<J}$, \revF{however the interaction should large enough ${\abs{U}>\Omega}$ such that the many-body effects appear.}
\revF{The off-resonant tunneling process at the ring-lead interface generates NOON-like states, with a superposition state of $N$ particles being in either upper or lower part of the ring.}
Well defined AB oscillations are found: the AB flux quantum decreases with particle number $N$ as ${\Phi_0=\frac{1}{N}}$.  \revF{A NOON state is a factor $N$ more sensitive to phase differences\cite{walther2004broglie}}. Thus, the fractional flux quantum in the interference pattern is the signature for the NOON-like state in the ring. At the AB minimum, one particle is reflected, while the rest is transmitted. The dependence on driving frequency is discussed in the supplementary materials.
The nature of the process and the fidelity of creating entangled state at the ring-lead junction is further discussed in the subsection Creating entangled states. %\revC{The fidelity at the ring-lead junction is discussed in the supplementary material.}
%For two distinguishable particles, we get a superposition of a Cooper-pair like bound state $\ket{\Psi_+}\propto \ket{\uparrow\downarrow| 0 }+\ket{0|\uparrow\downarrow }$.

{\bfseries Central band 0$^\pm$}
\begin{figure}[htbp]
	\centering
%	\subfigimg[width=0.24\textwidth]{a}{densityParam1DNplotFluxtimeFy0s1m12L8N4n0J1U0_09u0_0g1K1f0_0w0p60h0_75o-0_01s10000.pdf}\hfill
%	\subfigimg[width=0.24\textwidth]{b}{densityParam1DNplotFluxtimeFy0s1m12L8N4n0J1U-0_17u0_0g1K1f0_0w0p60h0_75o-0_01s10000.pdf}
\subfigimg[width=0.49\textwidth]{}{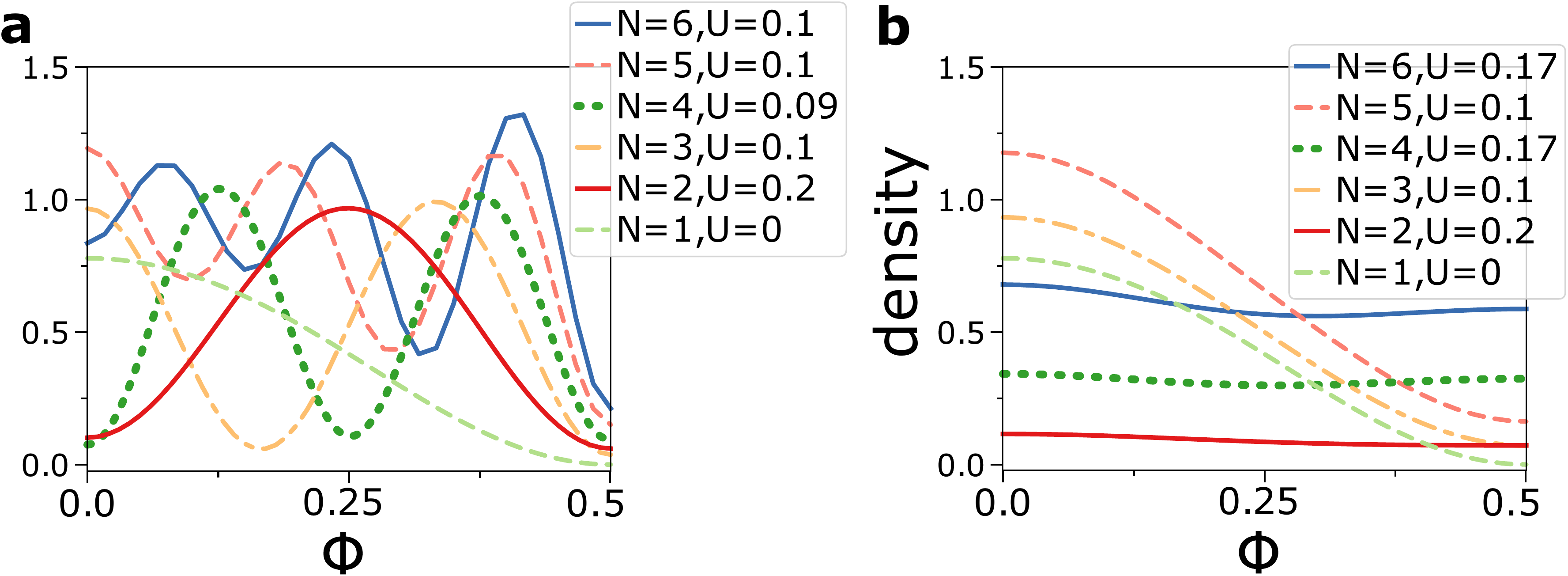}
\caption{{\it Transmission for central band} Density pumped into the drain for central band $0$ against flux for ${U>0}$. \idg{a} band $0^+$ ($\phi_0=\pi/2$): entangled state created by pumping results in fractional flux quantum $\Phi_0\propto 1/N$. \idg{b} $0^-$ ($\phi_0=-\pi/2$). Drain density measured at time ${tJ=630}$. Pumping with driving frequency ${\Omega=-0.01J}$, ring length $L_\text{R}=8$ and potential strength ${P_0=60J}$.}
	\label{Drainband0}
\end{figure}
%As the Chern number of the band is positive, the pumping occurs in opposite direction compared to band $\pm1$;  therefore, we change the sign of the driving ${\Omega\rightarrow-\Omega}$. 
%The particles are initialized at the first source site, directly neighboring the ring (see Fig.\ref{IntPumping}b). 
The dynamics for the central band  depends on the choice of initial phase: We define the initial conditions $0^+$  with ${\phi_0=\pi/2}$ and $0^-$ with ${\phi_0=-\pi/2}$. %(see  Fig.\ref{IntPumping}a). %Here we  change the sign of interaction $U$ {\blue{ WHY????}}.
%Changing the sign of interaction has the same effect as interchanging band $0^+$ and $0^-$.
The transmission depends on the ring length $L_\text{R}$ as well. 
For ${L_\text{R}=4n+2}$ and ${U>0}$, band $0^+$ has the same flux dependence as the upper band +1. For $0^-$ the flux dependence is the same as the lower band -1.
However, because of the different Chern number ${\mathcal{C}=2}$, in those two cases the particles move at twice the speed and in opposite direction (thus exchange ${\Omega\rightarrow-\Omega}$). The anti-crossing type alternates between top and bottom %(see Fig.\ref{IntPumping}a) 
and the smallest energy gap that limits the dynamics scales as in band +1 ($\Delta E\propto J^{N}/U^{N-1}$).

For ${L_\text{R}=4n}$, the dynamics is quite different. \revD{While in the previous case, nearly all particles were transmitted, now for ${L_\text{R}=4n}$ the system is highly reflective and we find that maximally one of $N$ incoming particles is transmitted.}
%We show this in Fig.\ref{Drainband0}. 
For the band $0^+$ (Fig.\ref{Drainband0}a) the AB flux quantum is ${\Phi_0=1/N}$, revealing the NOON-like state in the ring. The transmission is much lower and parity dependent: for even $N$ and zero flux, zero particles are transmitted, while for half flux quantum one particle is transmitted. For odd $N$, we find the opposite behavior: one particle transmitted at zero flux, and zero transmitted at half flux-quantum.
For band $0^-$ (Fig.\ref{Drainband0}b), the flux quantum is ${\Phi_0=1}$. For even $N$, the transmission is zero, while for odd $N$ it changes from one to zero with flux.
The dependence on $L_\text{R}$ comes from the switch between top and bottom approaches through anti-crossings  at every other site: For ${L_\text{R}=4n}$, the transitions at source-ring and ring-drain approach the avoided crossing from opposite ways; \revF{as the type of transition (resonant or off resonant) depends on the path in the anti-crossing,} this feature implies the  change in transmission behavior. For ${L_\text{R}=4n+2}$,  instead, the transitions at source-ring and ring-drain approach the avoided crossing the same way; this  features implies that the transmission is similar to what found for the band $\pm1$. \revF{The dependence on $\phi_0$ results also from switch between top and bottom approaches through anti-crossings: For ${\phi_0=\pi/2}$, the lead-ring junction is approached via the top path (off-resonant, NOON states), while for ${\phi_0=-\pi/2}$ via the bottom path (resonant).}
% in Fig.\ref{Centerband}.
%Transmission is close to zero for even number of particles. For odd, one particle is transmitted at zero flux and no particles at half flux-quantum.
In our numerical simulation, we see a finite probability of reaching the drain for even number of particles due to non-adiabatic transitions. The dependence on interaction is discussed in the supplementary materials.

%This causes the profound difference in the dynamics.

\begin{figure}[htbp]
	\centering
	\subfigimg[width=0.4\textwidth]{}{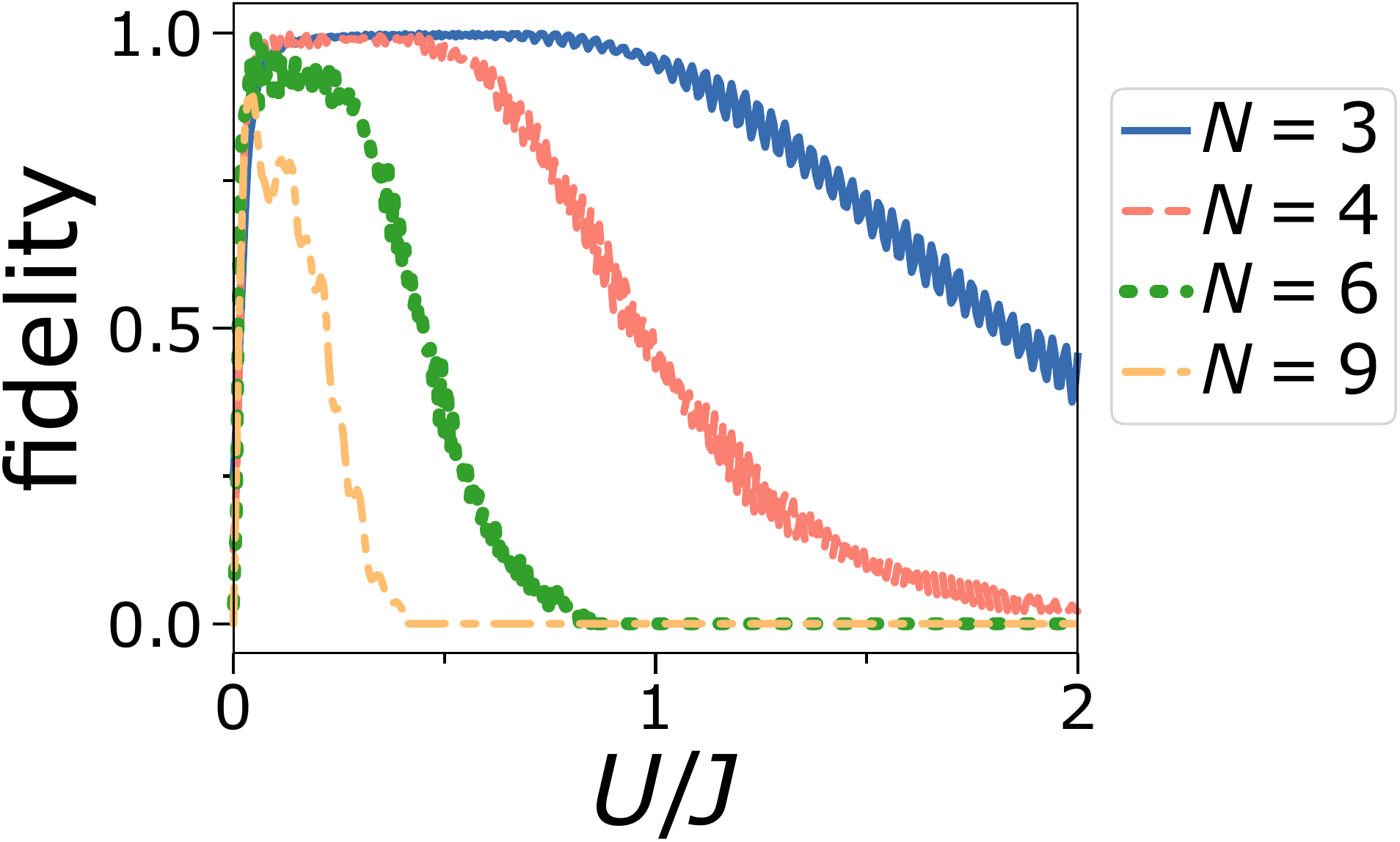}\hfill
	\caption{{\it Fidelity for NOON-state generation} Fidelity of creating a NOON-like entangled state after pumping through a simplified ring-lead junction. It consists of three lattice sites, with one site connected to the other two sites. All particles are initialized on the first site, the NOON state is measured between the other two sites. Fidelity is plotted against interaction $U$ in units of inter-site hopping $J$. Pumping with driving frequency ${\Omega=0.01J}$ and potential strength ${P_0=40J}$. }
	\label{NOON}
\end{figure}

{\bfseries Creating entangled states}
\revC{In this section we highlight how to create highly entangled states of NOON-type with our proposed setup. The crucial part to create entanglement is the source-ring junction, where one input site is connected to two output sites via tunneling. %(see Fig.\ref{IntPumping}c). 
This part represents a non-linear beam-splitter. To further understand the mechanism, we first look at a reduced system, where we remove all other sites of the ring-lead setup except this junction. We initialize the particles at the input site, and evolve the system.
Depending on $\phi_0$ of the potential, the state after the pumping can be highly entangled.
To create entangled states, we set the initial potential phase ${\phi_0=0}$ so that the system follows top curve of anti-crossing. %in Fig. \ref{IntPumping}b. 
The dynamics of this level is characterized by off-resonant coupling: the tunneling to the two neighboring sites is mediated by off-resonant energy levels as we explained in the subsection Upper band. Starting from the initial many-body state (for $N$ particles ${\ket{\Psi_\text{ini}}=\ket{N}\otimes\ket{00}}$) 
% as defined in Fig.\ref{IntPumping}c) 
the state is transformed directly to the final entangled state (${\ket{\Psi_\text{NOON}}=\frac{1}{\sqrt{2}}\ket{0}\otimes(\ket{N0}+\ket{0N})}$). %\rev{The tunneling happens when initial and final Fock states have about the same energy, and the particles tunnel together as a whole (due to the off-resonant transition), creating a highly entangled state in the process.}
We plot the fidelity ${F=\abs{\braket{\Psi_\text{NOON}}{\Psi}}^2}$ of the creation of the NOON like entangled state by adiabatically \revF{changing} the potential in the ring-lead junction in Fig.\ref{NOON}. We observe that for our parameters a NOON state of up to 6 particles with nearly unit fidelity can be created.} \rev{For more particles or higher interaction the fidelity decreases due to the exponential suppression of the energy gap.}

\revC{In the other case, where the particles follow the lower band of the anti-crossing with ${\phi_0=\pi}$, different types of states can be created. The tunneling from one site to the next occurs via resonant tunneling between the intermediate many-body states. Particles tunnel one after the other over to the neighboring sites when the states are brought into resonance by the driving. For ${N=2}$ %and same notation as in Fig.\ref{IntPumping}c
, the initial state $\ket{2}\otimes\ket{00}$ transforms to final state $\ket{0}\otimes\ket{11}$ via resonantly occupying the intermediate states ${\ket{1}\otimes(\ket{10}+\ket{01})}$.}] %Further possible states are shown in Fig.\ref{IntPumping}c.}
Driving this setup with two species of interacting particles (e.g. spin up and down) with an initial state $\ket{\Psi_0=\ket{\uparrow\downarrow}}\otimes\ket{00}$, a Bell state can be created: ${\ket{\Psi_\text{Bell}}\propto \ket{0}\otimes(\ket{\uparrow|\downarrow}+\ket{\downarrow|\uparrow})}$ (see also the supplementary materials).

\section{Discussion}
\revG{We studied topological pumping in an interacting ring-lead system pierced by a synthetic magnetic field.}
\revG{Due to the interplay between topological bands and AB phase we find that the transport is substantially affected by entanglement and interaction.}
%We find that the interplay between topological bands and Aharonov-Bohm phase produces {\it entangled states} a complex variety of flux quanta, types of entangled states and transmission coefficients. 

\revD{Interaction fundamentally influences the topological pumping, giving each band distinct dynamics and characteristics: The lowest band has the largest energy gap. The central band either transmits or reflects nearly all incoming particles depending on the parity of the ring length.}
\revC{The top band creates highly entangled states.}

\revG{Entanglement is generated because the ring-lead interface effectively acts as a nonlinear (Bose-Hubbard interaction) beam splitter. The oscillations of the particle transmission then occurs with a specific periodicity due to the AB interferometer. Such phenomenon traces back to the entangled nature (NOON-type) of the states involved in the transport, thus forming a fractional flux quantum.
\revC{This effect may be used for quantum-enhanced interferometers with a sensitivity ${\propto1/N}$. We note that the state is prepared adiabatically.
Adiabatic shortcuts could \revG{reduce their preparation time}\cite{torrontegui2013shortcuts}. 
In addition, these entangled states are limited by an energy gap that scales exponentially with the number of particles $N$. }\rev{This process is distinct from the Hong-Ou-Mandel effect which is based on two-photon interference\cite{hong1987measurement}.}
\revG{We note that in the lowest band, the energy gap scales as ${\Delta E\propto\sqrt{N}}$, but the state involved in the transport is not of NOON-type.}}
%\revD{It is of direct experimental interest for various platforms of quantum technologies. Topological pumping can be robust to disorder and particle loss \cite{niu1984quantised,tangpanitanon2016topological,ke2017multiparticle} which eases experimental realization. Noise in general has only minor influence on the adiabatic dynamics\cite{fubini2007robustness}.
	%Cold atoms smooth traps etc. boshier, aghamalayan, TAPP klitzing

%The topological pumping scheme would make this device robust to disorder and imperfections.
%\revF{Disorder would impact the interferometry, as disorder at the ring-lead interface can create a (random) particle imbalance and phase difference between particles moving the two paths in the ring. \revG{However, it is reasonable to assume that the charge transport of particles itself is robust against disorder as long as the disorder is less than the energy gap.}}
\revG{Disorder at the ring-lead interface can create a (random) particle imbalance as well as a phase shift between particles in the upper and lower part of the ring. When disorder is applied away from the ring-lead interface small disorder  (compared with the energy gap) is expected not to harm the pumping protocol. We believe that a separate study should be carried out to analyse this problem.}

%Our results could be used to engineer a wide variety of entangled states for quantum technology application, like highly entangled NOON states\cite{afek2010high}.
%Relying on  the established techniques available in quantum technology our results are of immediate interest for cold atoms experiments, superconducting resonators and photonic waveguides.
\revG{Our setup is of direct experimental interest for various platforms of quantum technologies.}
\revD{For cold atom systems, specifically,  decoherence is well controlled as well as smooth traps  can be engineered to limit disorder\cite{navez2016matter,Boshier_integrated,amico2014superfluid}.}
Interacting photons in non-linear superconducting resonators can realize topological pumping of interacting photons%, which is robust to loss of excitations 
\cite{tangpanitanon2016topological} and synthetic magnetic fields \cite{roushan2017chiral}.
Photonic waveguides could realize flexible designs of topological pumping\cite{zilberberg2018photonic} and artificial magnetic fields\cite{mukherjee2018experimental}, while interaction between photons can be engineered via strong coupling\cite{hartmann2016quantum}. 
%. In these systems, topological pumping of interacting photons \cite{tangpanitanon2016topological} and synthetic gauge fields \cite{roushan2017chiral} has been demonstrated.

%Our setup opens up a path to generate highly entangled NOON states in the upper and lower part of the ring, which from there could be used for quantum information purposes. 
%The speed of preparation is limited by the condition of adiabatically for the driving frequency and has to become exponentially small with particle number $N$ as $\log(\Omega/J)\ll(N-1)\log(U/J)$. 

\section{Methods}
The equations of motions of the system are solved with exact diagonalization, by evolving the Schr\"odinger equation in time ${\ket{\Psi(t)}=\expU{-i\mathcal{H}t}\ket{\Psi(0)}}$.

Transmission and reflection coefficients are derived by evaluating the dynamics of the pumped eigenstates of the reduced system: We consider only the bare junction, consisting only of three sites: one input site, and two output sites. To get the transmission and reflections coefficients, the dynamics of source and drain junction are considered.

\begin{acknowledgments}
%{\it Acknowledgments}
\section{Acknowledgments} We thank V. Bastidas, A. Daley, W. Munro and J. Tangpanitanon for discussions. The Grenoble LANEF framework (ANR-10-LABX-51-01) is acknowledged for its support with mutualized infrastructure. We thank National Research Foundation Singapore and the Ministry of Education Singapore Academic Research Fund Tier 2 (Grant No. MOE2015-T2-1-101) for support. The computational work for this article was partially performed on resources of the National Supercomputing Centre, Singapore (https://www.nscc.sg).
\end{acknowledgments}

\bibliography{library}

\appendix
\section{Scaling of the energy gap}
\begin{figure}[htbp]
	\centering
	\subfigimg[width=0.24\textwidth]{a}{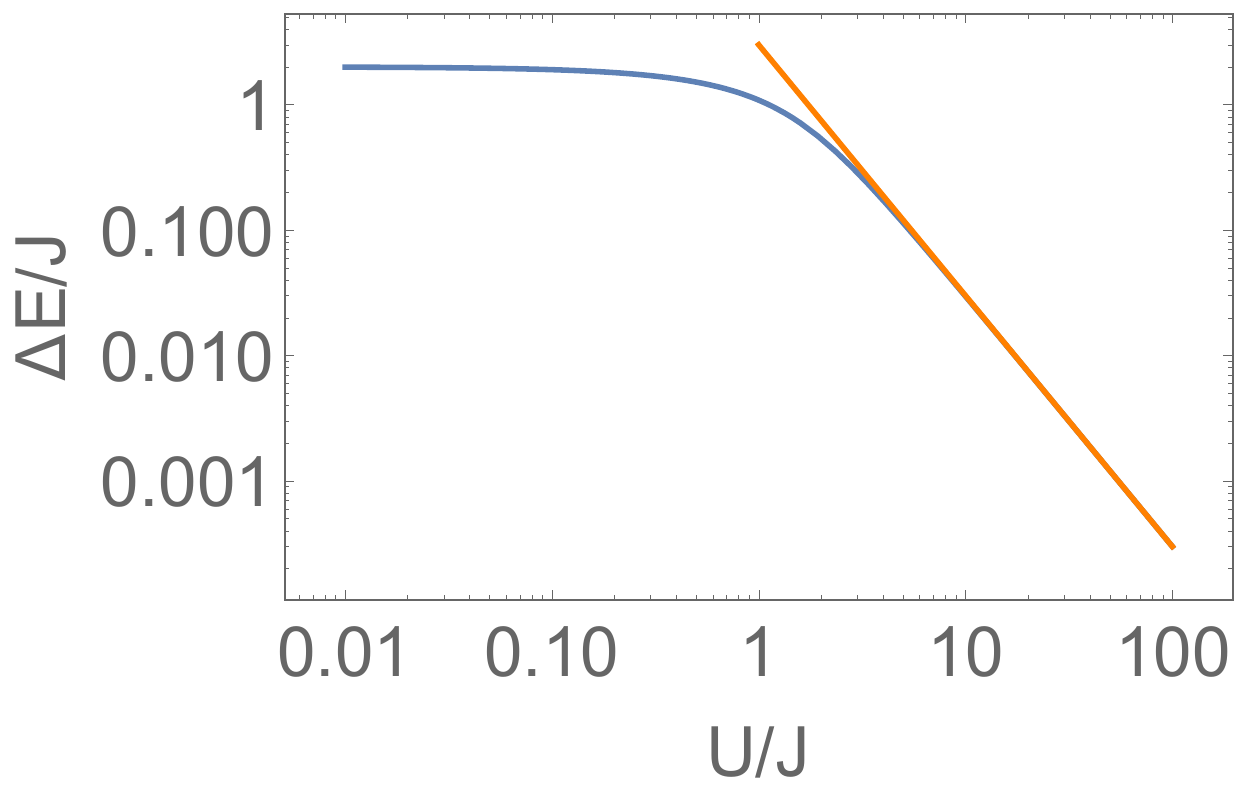}\hfill
	\subfigimg[width=0.24\textwidth]{b}{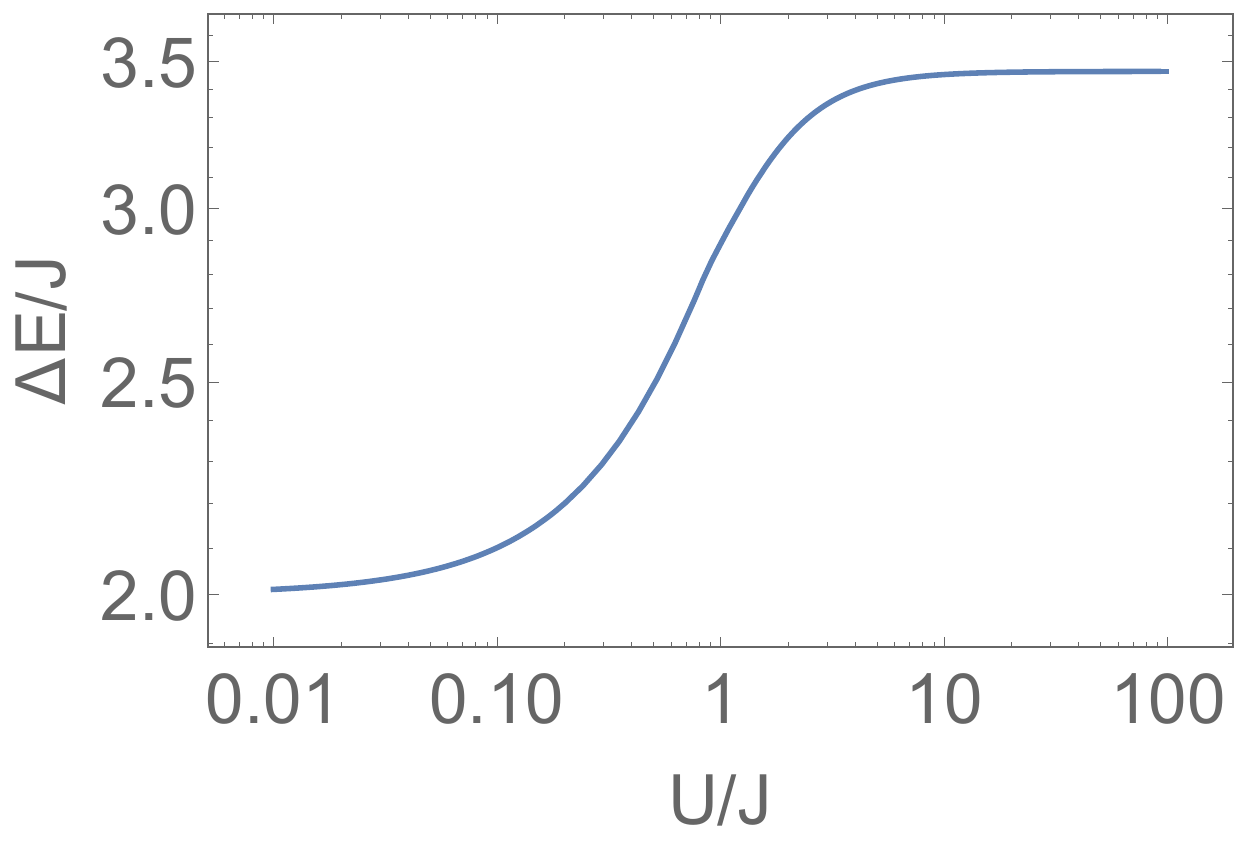}
	\caption{ Scaling of the energy gap $\Delta E$ at the anti-crossing. Energy gap in blue, fit in orange. \idg{a} Off-resonant coupling (${U>0}$, top curve of anti-crossing as in band +1). For small $U$, gap scales as ${\Delta E\propto J^N/U^{N-1}}$, indicated in orange (here ${N=3}$). To avoid non-adiabatic Landau-Zener transitions, the driving speed must be $\Omega\ll\Delta E$. Gap is largest for $U<J$.
		\idg{b} Scaling of the energy gap $\Delta E$ for resonant transitions (${U>0}$, bottom curve of anti-crossing as in band -1). Energy gap ${\Delta E=2\sqrt{N}J}$  is independent of $U$ for ${\abs{U}\gg J}$. }
	\label{Scaling}
\end{figure}

The energy gap between the pumped many-body energy level and the next level determines the rate of non-adiabatic Landau-Zener transitions. To avoid these transitions the driving frequency $\Omega$ must be smaller than the energy gap ${\Omega\ll\Delta E}$. In Fig.\ref{Scaling}, we plot the lowest energy gap to the next level for top and bottom anti-crossing curves. 
In the following, we derive the many-body energy gap for bottom curve. We consider a reduced model, consisting of two neighboring sites only. Other sites can be neglected due to the large potential difference. We indicate the many-body states of this two-site model as $\ket{N_1,N_2}$, where $N_1$ ($N_2$) is the number of particles at the first (second) site. For bottom curve of anti-crossing, the many-body states are resonantly coupled. As a result, for e.g. ${N=3}$, the following resonant transitions occur one after the other due to the adiabatic driving: ${\ket{30}\rightarrow\ket{21}\rightarrow\ket{12}\rightarrow\ket{03}}$. For ${\abs{U}\gg J}$, the minimal energy gap is given by ${\Delta E=2\sqrt{N}J}$. It can be derived by degenerate perturbation theory (for ${P_0\gg U\gg J}$). At a specific point in time, two many-body states will have the same local energy (potential plus interaction energy). All other many-body states are far off-resonant. The two resonant states are weakly coupled by the hopping strength $J$. We now give an example for two Fock states $\ket{N,0}$ and $\ket{N-1,1}$ and ${U>0}$: The local energy of state $\ket{N,0}$ is $E_1=-NP_0\cos\left(\Omega t\right)+\frac{U}{2}N(N-1)$, and of $\ket{N-1,1}$ $E_2=-(N-1)P_0\cos\left(\Omega t\right)-P_0\cos\left(2\pi/3+\Omega t\right)+\frac{U}{2}(N-1)(N-2)$. At a time $t^*$, both levels are degenerate $E_1(t^*)=E_2(t^*)$ with all other many-body states far off-resonant. We can treat them as a two level system within degenerate perturbation theory. The two states are now weakly coupled via the nearest-neighbor hopping $\mathcal{H}_J=-J\cn{a}{1}\an{a}{2} +\text{H.C.}$. The coupling matrix element is $\bra{N,0}\mathcal{H}_J\ket{N-1,1}=\sqrt{N}J$. The resulting energy splitting is ${\Delta E=E_2-E_1=2\sqrt{N}J}$. This is the minimal energy gap, all further resonant transitions with other Fock states have larger gaps.

\section{Frequency dependence}
\begin{figure}[htbp]
	\centering
	\subfigimg[width=0.24\textwidth]{a}{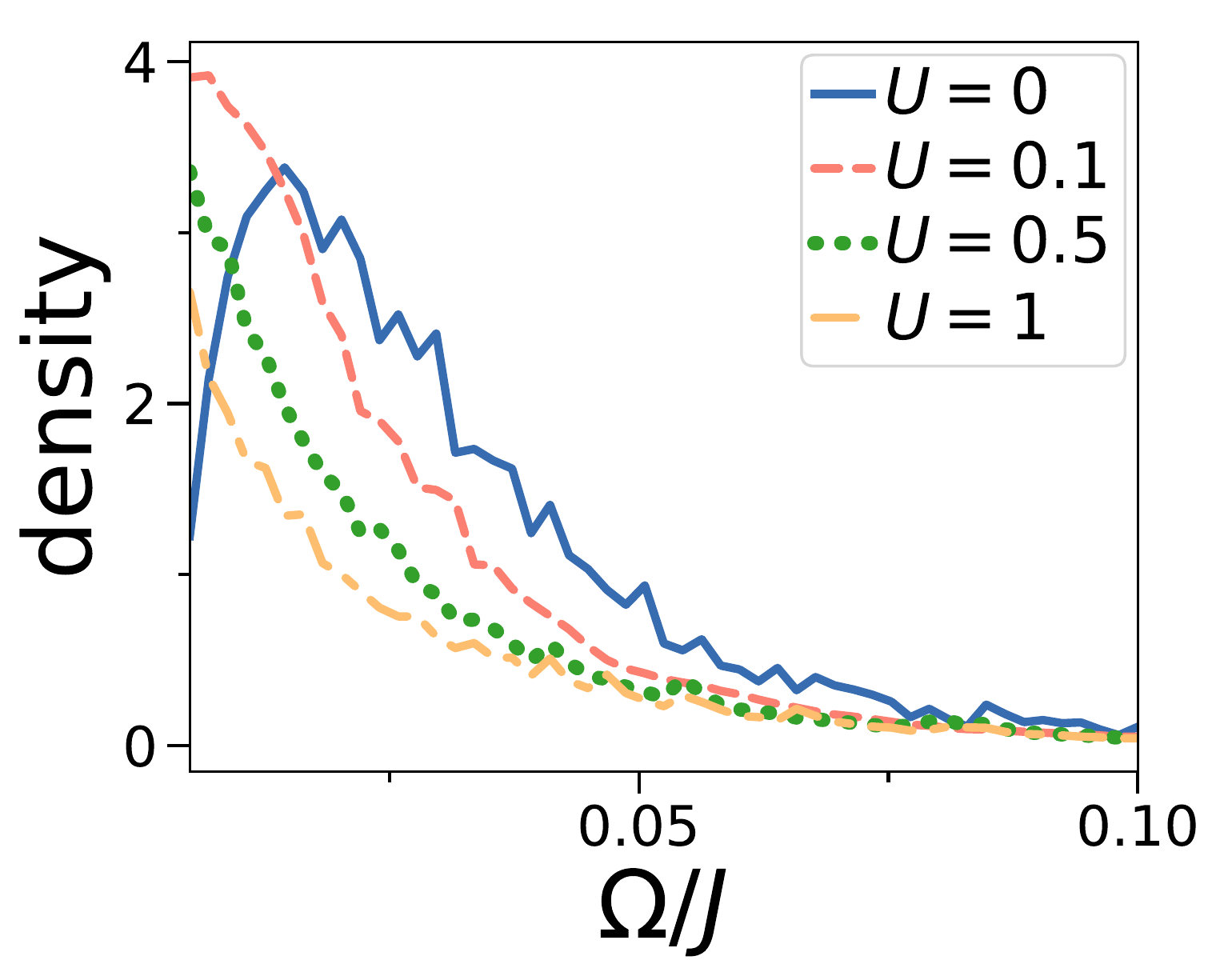}\hfill
	\subfigimg[width=0.24\textwidth]{b}{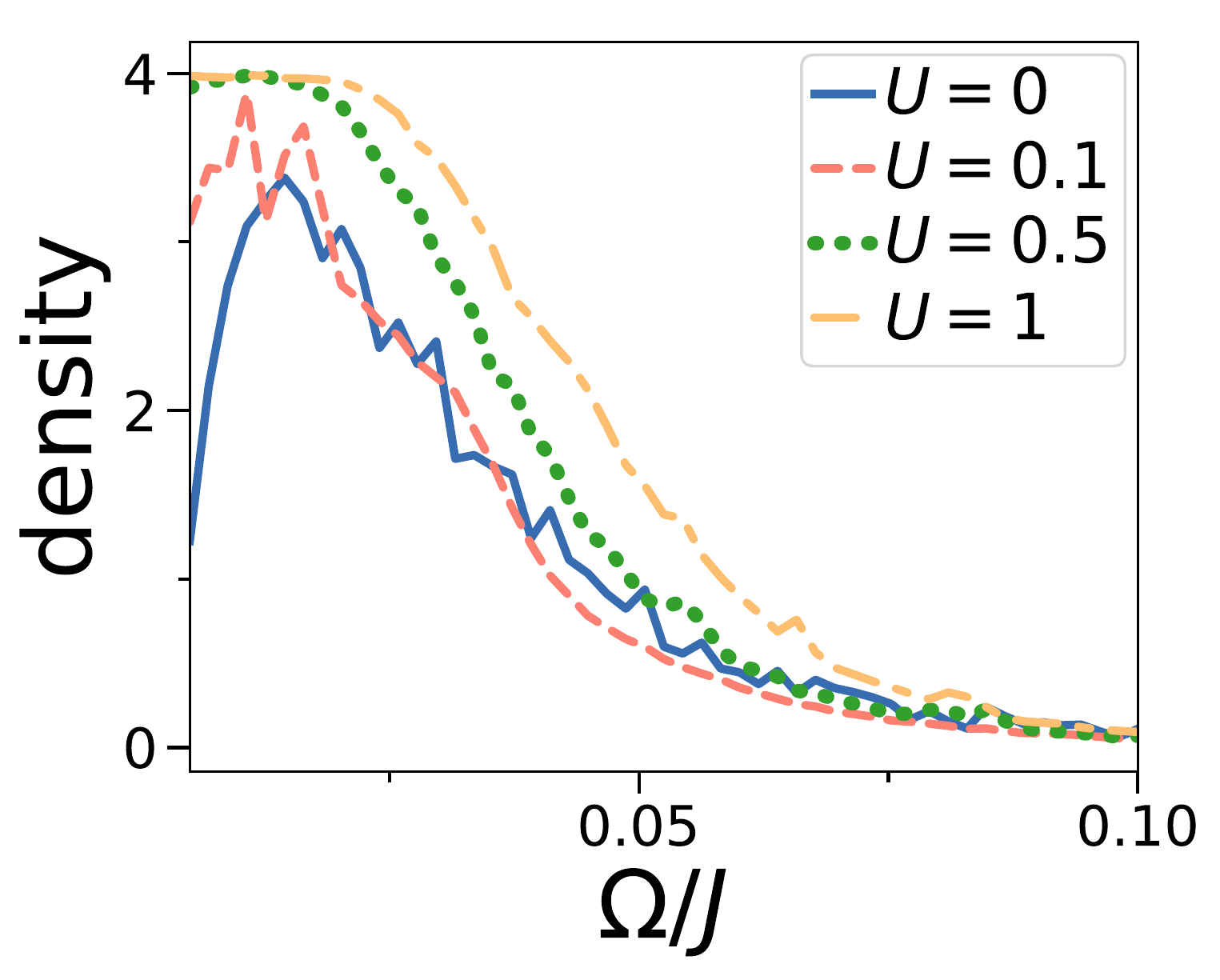}
	\caption{Density pumped through the ring into the drain plotted against driving frequency $\Omega$ for \idg{a} top band +1 and \idg{b} bottom band -1.
		Drain density taken at ${t=12.6/\Omega}$ with ring length ${L_\text{R}=8}$ and potential strength ${P_0=60J}$ and ${N=4}$ particles. }
	\label{DrainDensFreq}
\end{figure}

We show the frequency dependence of the pumping protocol for different values of interaction and bands in Fig.\ref{DrainDensFreq}. We observe that for bottom band in Fig.\ref{DrainDensFreq}b, fidelity increases with interaction. For top band in Fig.\ref{DrainDensFreq}a, the fidelity decreases for large interaction. For zero interaction, we observe a more complex behavior: For slow driving, fidelity is very low and then increases and then decreases again for higher driving.  The reason for that is effective three site tunneling: We initialize atoms at a single site, however or potential has a period of three sites. Particles can tunnel by three sites to a site with degenerate potential energy. This effect comes into play only when driving is very slow. With interaction $U$, this effective three site tunneling is suppressed.

\section{Interaction dependence}
\begin{figure}[htbp]
	\centering
	%\subfigimg[width=0.24\textwidth]{a}{densityParam1DUtimeFy0s1m12L8N2n0J1U-0_5u0_0g1K1f0_0w0p60o0_01s20000.pdf}\hfill
	%\subfigimg[width=0.24\textwidth]{b}{densityParam1DUinttimeFy0s1m12L6N1n1J1U0_0u-2_0g1K1f0_0w0p60o0_01s10000n.pdf}\\
	\subfigimg[width=0.24\textwidth]{a}{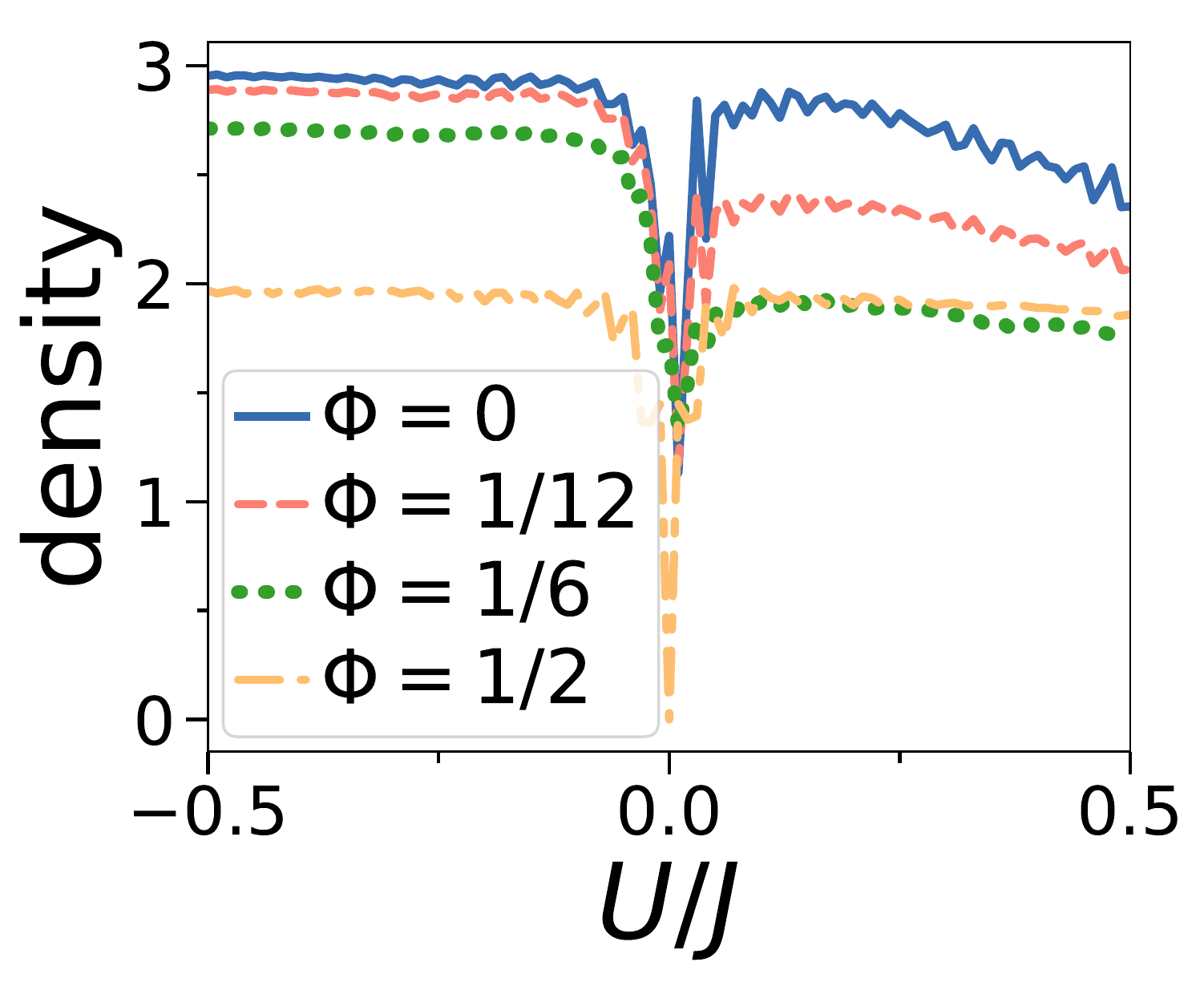}\hfill
	\subfigimg[width=0.24\textwidth]{b}{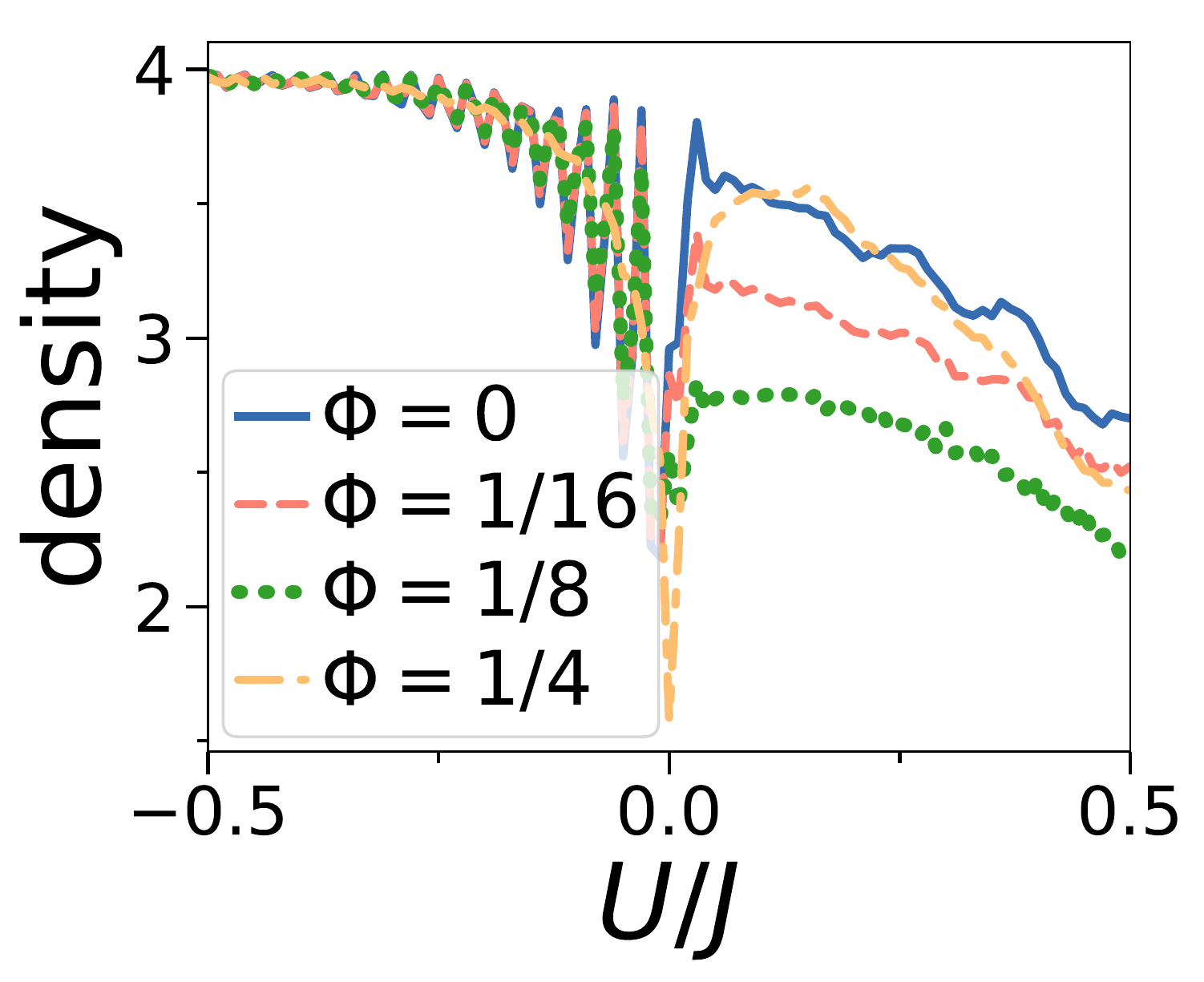}
	\caption{Density pumped through the ring into the drain plotted against interaction strength for band +1 (For dynamics of band -1, invert interaction ${U\rightarrow-U}$).
		%\idg{b} 2 distinguishable particles 
		\idg{a} 3 particles: for ${U<0}$ AB effect changes particle number in drain between 3 and 2, with minimum for ${\Phi=1/2}$. For ${U>0}$ same change in density, however with minimum at ${\Phi=1/6}$.
		\idg{b} 4 particles: for ${U<0}$ independent of flux, while for ${U>0}$ minimum at ${\Phi=1/8}$. Close to ${U\approx0}$, minimal density for ${\Phi=1/2}$.  The number of particle pumped into the drain decreases for larger $U$ due to Landau-Zener transitions as the gap decreases, while for negative interactions it does not change with large negative interactions. The strong oscillations at small $U$ result from non-adiabatic transitions when $U\approx\Omega$.
		Drain density taken at ${tJ=1260}$ with ${L_\text{R}=8}$, ${\Omega=0.01J}$ and ${P_0=60J}$. }
	\label{DrainDensInt}
\end{figure}

\begin{figure}[htbp]
	\centering
	%\subfigimg[width=0.24\textwidth]{a}{densityParam1DFluxtimeFy0s1m12L8N3n0J1U0_1u0_0g1K1f0_0w0p60h0_75o-0_01s10000n.pdf}\hfill
	%\subfigimg[width=0.24\textwidth]{b}{densityParam1DFluxtimeFy0s1m12L8N4n0J1U0_09u0_0g1K1f0_0w0p60h0_75o-0_01s10000.pdf}\\
	\subfigimg[width=0.24\textwidth]{a}{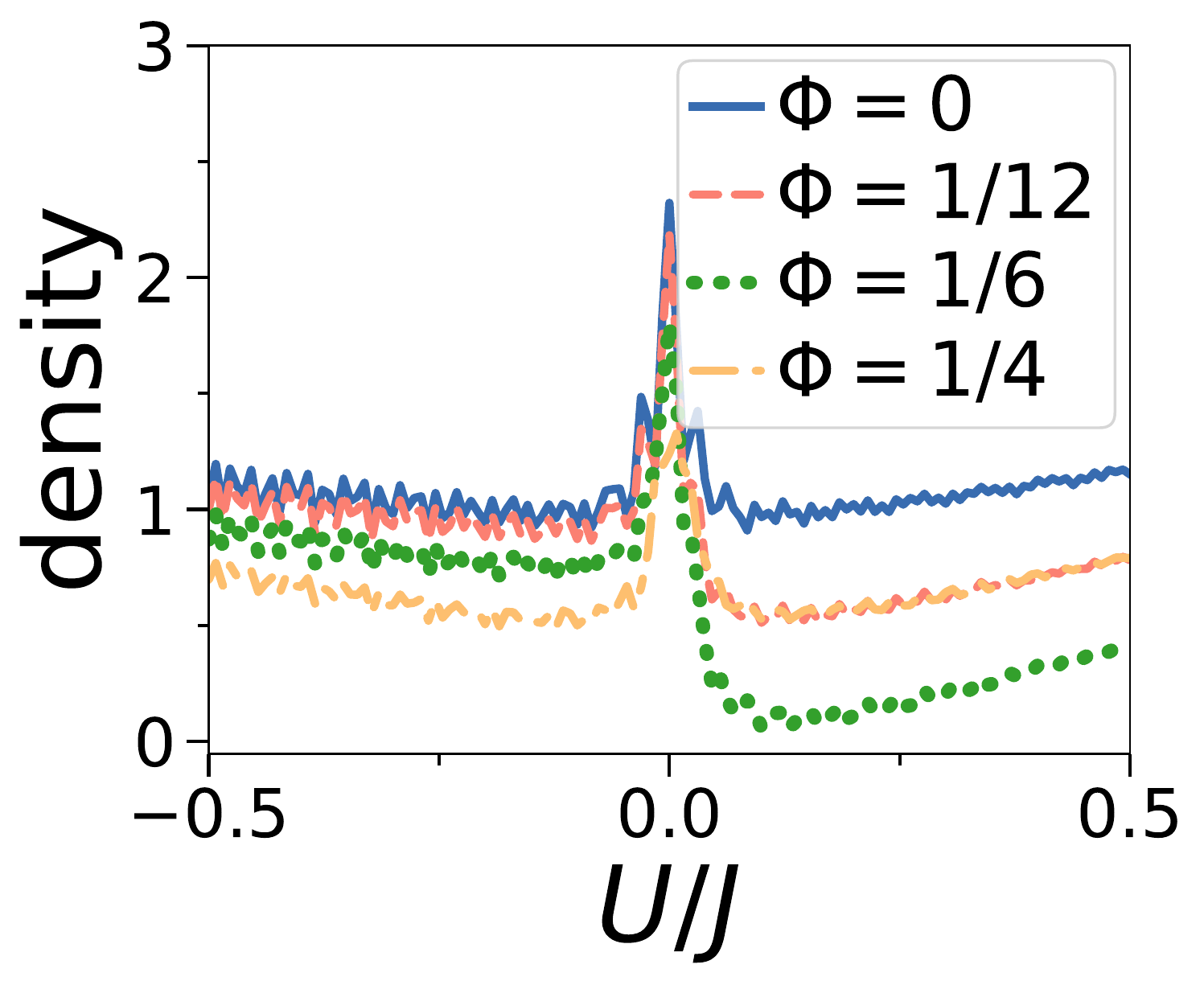}\hfill
	\subfigimg[width=0.24\textwidth]{b}{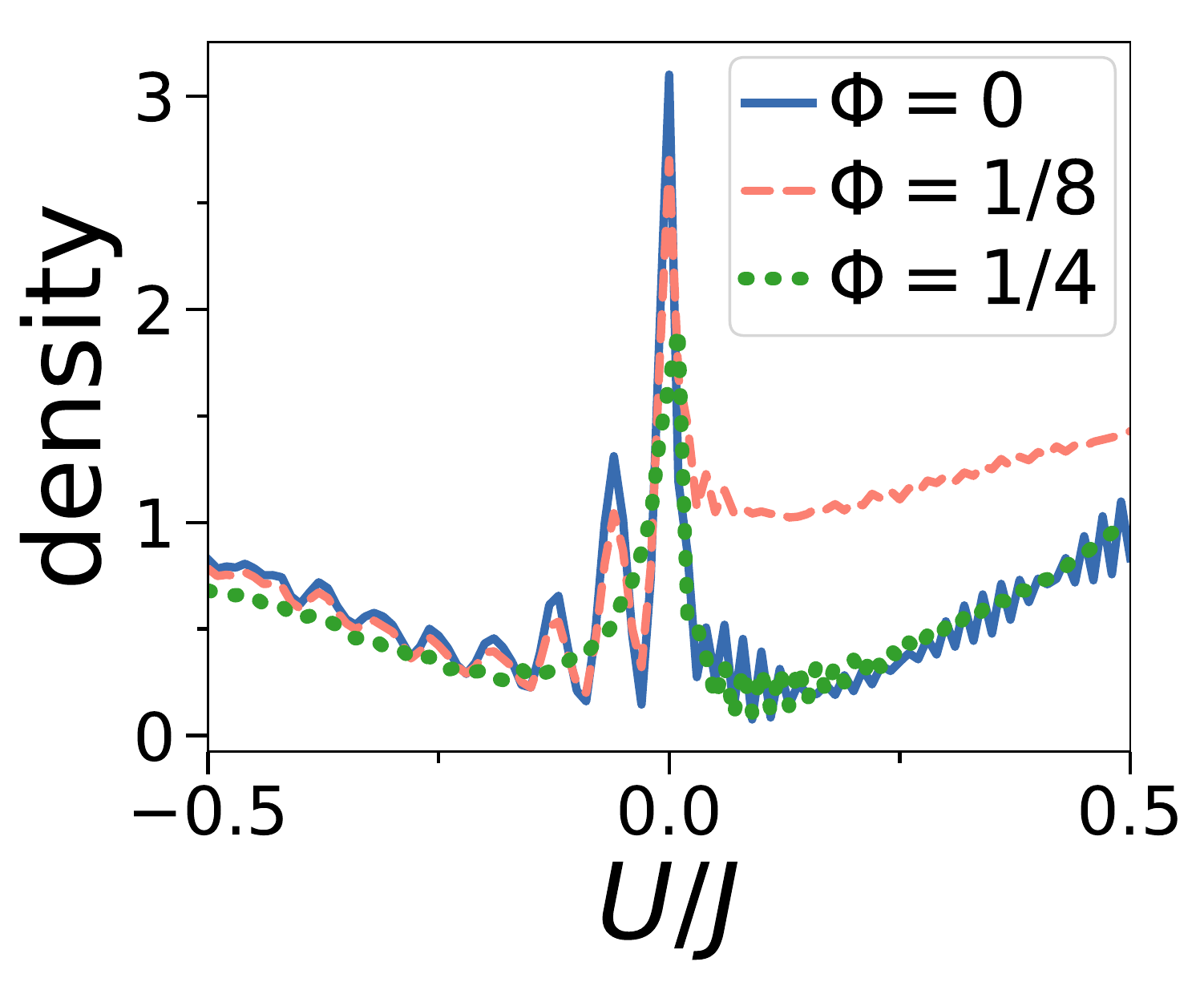}
	\caption{Transmitted density pumped into the drain for central band with Chern number ${\mathcal{C}=2}$ . Here, band 0$^+$ with initial condition ${\phi_0=\pi/2}$ was chosen. (For dynamics of band 0$^-$, invert interaction ${U\rightarrow-U}$) 
		%\idg{a} Flux dependence for ${N=3}$ particles and  
		%\idg{b} ${N=4}$ particles. 
		\idg{a} Dependence on interaction $U$ for different values of flux $\Phi$ for ${N=3}$ particles and 
		\idg{b} for ${N=4}$ particles. 
		Drain density taken at ${tJ=630}$ with ${L_\text{R}=8}$, ${\Omega=-0.01J}$ and ${P_0=60J}$.   }
	\label{Centerband}
\end{figure}

In this section we study the transmission into the drain for varying interaction at different values of flux and particle number. For band +1 and -1, see Fig.\ref{DrainDensInt}. For the central band 0, see Fig.\ref{Centerband}.

\section{Two species pumping}\label{DistingiPart}
One can also consider a different setup: The same Hamiltonian as introduced in the main text with now two species of particles (e.g. two internal states, denoted as $\uparrow$, $\downarrow$). There is only one particle of each species. The two species interact with 
\begin{equation}
\mathcal{H}_\text{int}=\sum_j U\nn{\uparrow}{j}\nn{\downarrow}{j}
\end{equation} 
Driving this setup for the lower band -1 generates Bell states in the ring  ${\ket{\Psi_+}\propto \ket{\uparrow|\downarrow}+\ket{\downarrow|\uparrow}}$ (first part of bracket indicates type of particle in upper part of ring, and second part of bracket type of particle in lower part of ring). The pumping is flux  independent.
For the upper band +1 we observe entangled states.

\begin{figure}[htbp]
	\centering
	\subfigimg[width=0.24\textwidth]{a}{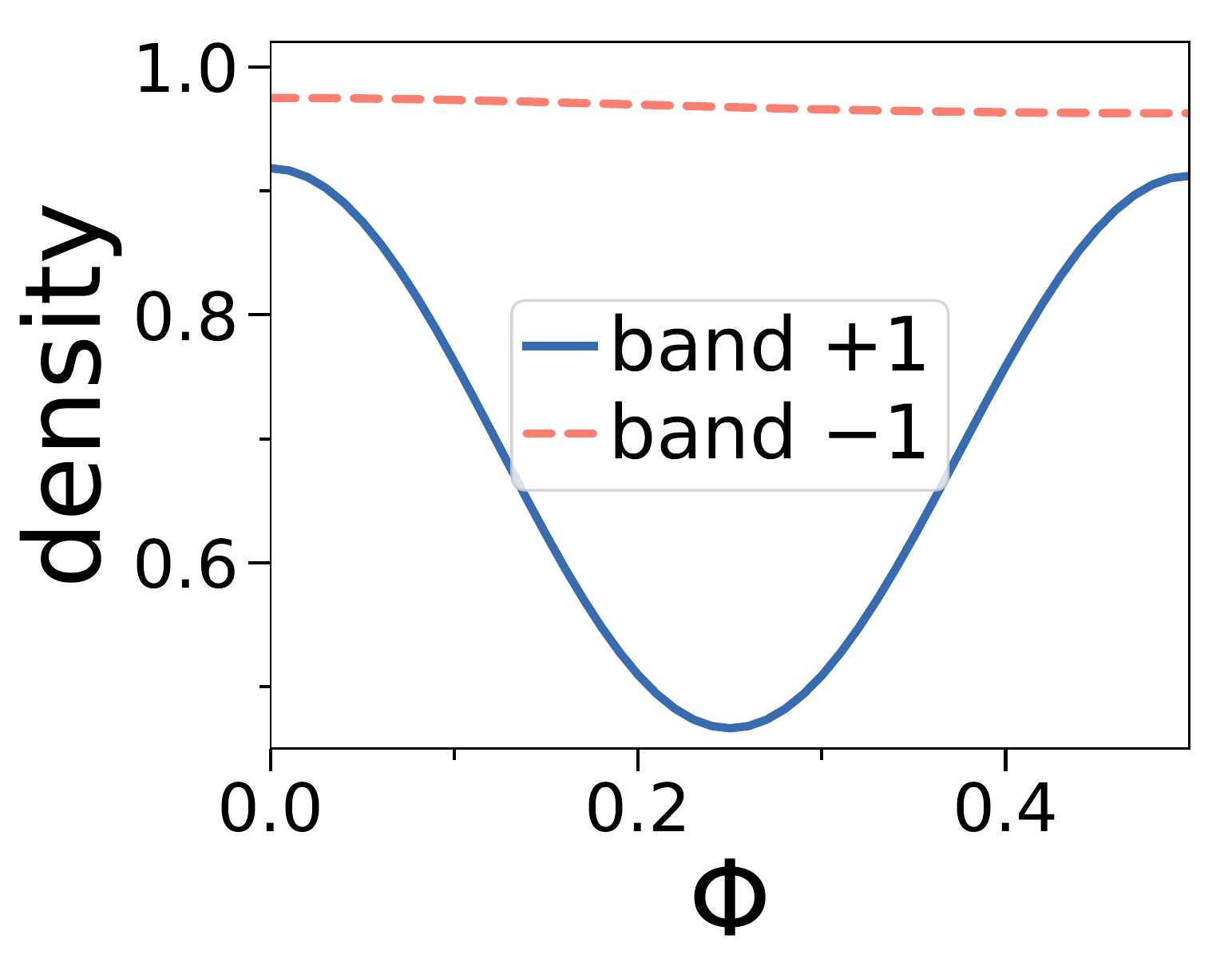}\hfill
	\subfigimg[width=0.24\textwidth]{b}{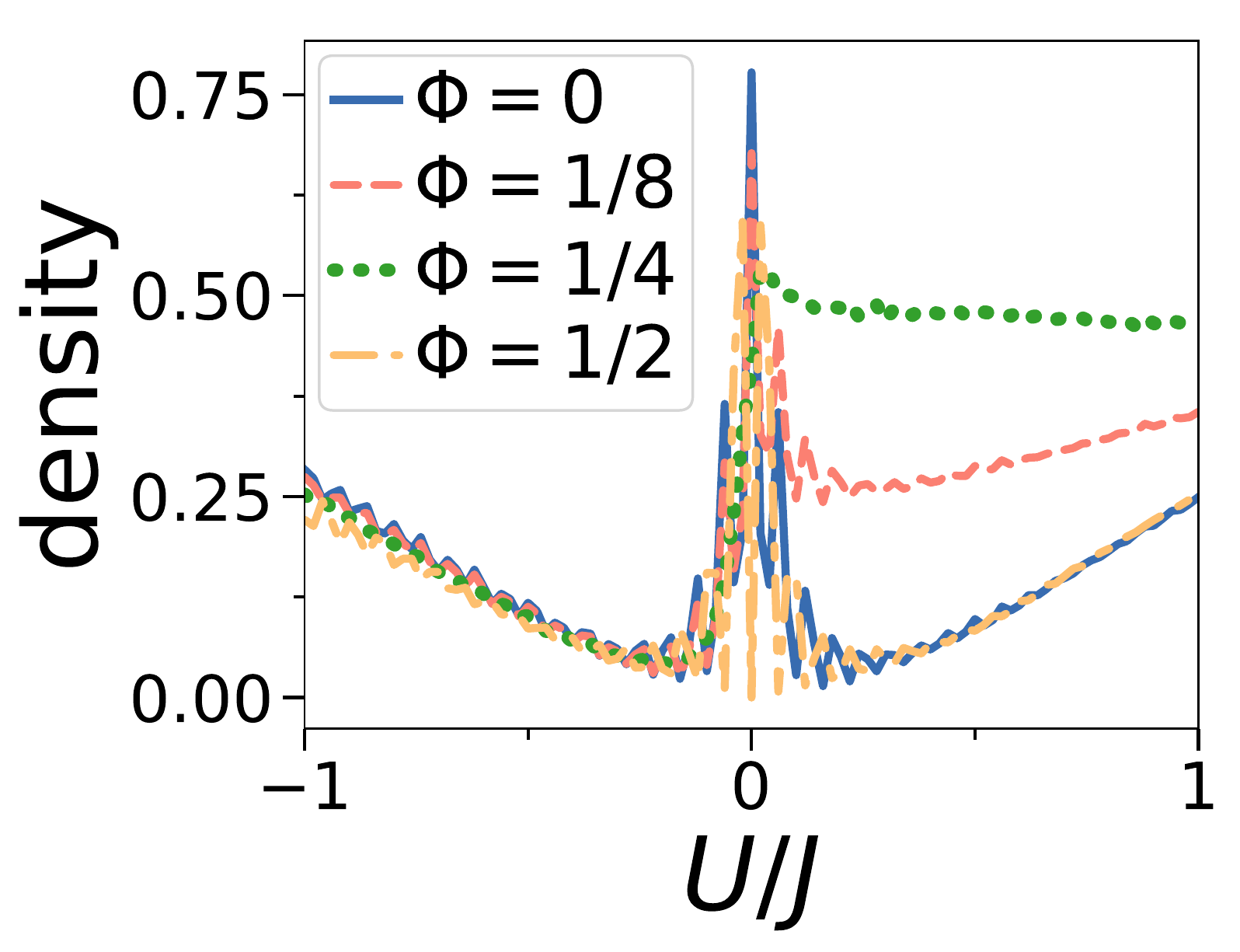}
	\caption{Density pumped through the ring into the drain for 2 types of particles, with one particles for each type. \idg{a} ${L=8}$, ${U=0.5}$ \idg{b} interaction dependence for central band with ring length ${L_\text{R}=8}$ and initial condition 0$^+$.  }
	\label{Disting}
\end{figure}

In Fig.\ref{Disting}, we present the transmitted density for two species of particles.

\end{document}